\documentclass[12pt]{paper}

\pdfoutput=1

\usepackage{amsfonts,amsmath,amssymb,amsthm}
\usepackage{epic,eepic}
\usepackage{graphics}
\usepackage{graphicx}
\usepackage{subfig}
\usepackage{multirow}

\usepackage{jneurosci}

\newcommand{\beq}{\begin{equation}}
\newcommand{\eeq}{\end{equation}}
\newcommand{\beqr}{\begin{eqnarray}}
\newcommand{\eeqr}{\end{eqnarray}}
\newcommand{\beqrn}{\begin{eqnarray*}}
\newcommand{\eeqrn}{\end{eqnarray*}}
\newcommand{\beqn}{\begin{equation*}}
\newcommand{\eeqn}{\end{equation*}}
\newcommand{\bei}{\begin{itemize}}
\newcommand{\beii}{\begin{itemize} \item}
\newcommand{\eei}{\end{itemize}}
\newcommand{\ben}{\begin{enumerate}}
\newcommand{\een}{\end{enumerate}}
\newcommand{\bes}{\begin{small}}
\newcommand{\ees}{\end{small}}
\newcommand{\bec}{\begin{center}}
\newcommand{\eec}{\end{center}}

\newcommand{\sig}{\sigma}
\newcommand{\lsim}{\mathrel{\hbox{\rlap{\lower.55ex \hbox{$\sim$}} \kern-.3em \raise.4ex \hbox{$<$}}}}
\newcommand{\gsim}{\mathrel{\hbox{\rlap{\lower.55ex \hbox{$\sim$}} \kern-.3em \raise.4ex \hbox{$>$}}}}

\newcommand{\Cov}{\mathrm{Cov}}
\newcommand{\Var}{\mathrm{Var}}
\newcommand{\Ex}{\mathop{\bf E\/}}
\newcommand{\STA}{\mathrm{STA}}

\begin{document}

\bibliographystyle{jneurosci}

\title{The A-current and Type I / Type II transition determine collective spiking from common input}

\author{ \large{Andrea K. Barreiro, Evan L. Thilo, Eric Shea-Brown$^*$} \\  \\\small{Department of Applied Mathematics and}\\ \small{Program in Neurobiology and Behavior}  \\ \small{University of Washington,}
\\ \small{Box 352420, Seattle, WA 98195} \\
\small{$^*$ Correspondence:  etsb@uw.edu }}

\maketitle

%\paragraph{Abbreviated title:}  Correlation transfer and neural excitability
%
%\bigskip
%
%\noindent{75 pages, 10 figures, 1 table} 
%
%\bigskip
%
%\noindent{Abstract:  250 words; Introduction: 520 words; Discussion:  1366 words.} 
%

\paragraph{Acknowledgements:}
We thank Kresimir Josi\'{c} and Brent Doiron for their helpful comments as this work progressed.  This research was supported by NSF grants DMS-0817649 and CAREER DMS-1056125, and by a Career Award at the Scientific Interface from the Burroughs-Wellcome Fund (ESB), by the Mary Gates Foundation at the University of Washington (ET), and by NSF Teragrid allocation TG-IBN090004.

\newpage

\section*{ABSTRACT}

The mechanisms and impact of correlated, or synchronous, firing among pairs and groups of neurons is under intense investigation throughout the nervous system. A ubiquitous circuit feature that can give rise to such correlations consists of overlapping, or common, inputs to pairs and populations of cells, leading to common spike train responses.  Here, we use computational tools to study how the transfer of common input currents into common spike outputs is modulated by the physiology of the recipient cells.  We focus on a key conductance --- $g_A$, for the A-type potassium current --- which drives neurons between ``Type II" excitability (low $g_A$), and ``Type I" excitability (high $g_A$).

% as for pyramidal cells.  Specifically, we ask how this transition affects the collective spiking of neurons driven with common fluctuating inputs.

Regardless of $g_A$, cells transform common input fluctuations into a tendency to spike nearly simultaneously.  However, this process is more pronounced at low $g_A$ values, as previously predicted by  reduced ``phase" models.
% that describe cells in mean-driven firing states.  
Thus, for a given level of common input, Type II neurons produce spikes that are relatively {\it more} correlated over short time scales.   
Over long time scales, the trend reverses, with Type II neurons producing relatively {\it less} correlated spike trains.  This is because these cells' increased tendency for simultaneous spiking is balanced by opposing tendencies at larger time lags.  We demonstrate a novel implication for neural signal processing:  downstream cells with long time constants are selectively driven by Type I cell populations upstream, and those with short time constants by Type II cell populations.  Our results are established via high-throughput numerical simulations, and explained via the cells' filtering properties and  nonlinear dynamics.

%, by computing the resulting spike-count correlations on different time scales.  {\bf Our methods combine high-throughput sims, STA-based analysis, bifurcations ...  }

\section*{INTRODUCTION}

%%%%%%%%%%%%%%%%%%%%%%%%%%%%%%%%%%%%%%%%%%%%%%%
 \begin{figure}[t]
 \vspace{-.5cm}
     \begin{center}
 {\includegraphics[width=4.5in]{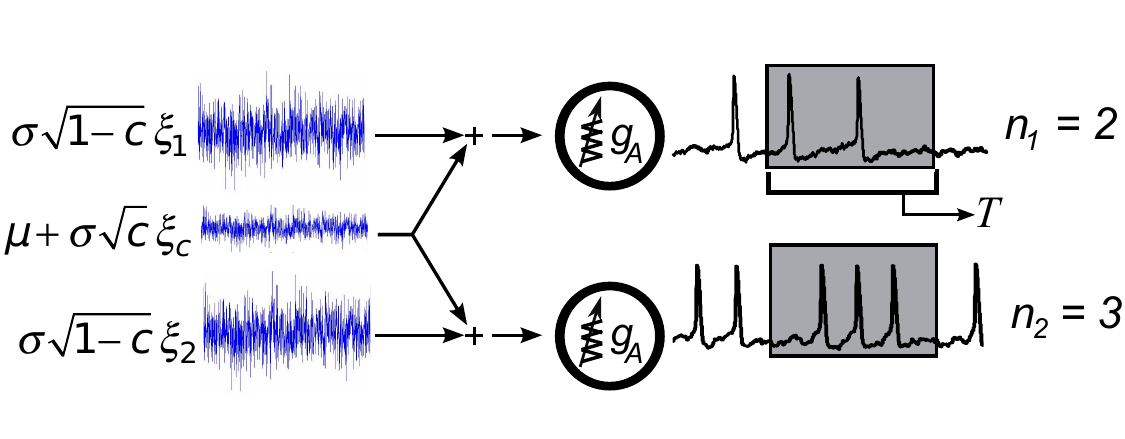}}  %\vspace{-.5cm}
     \end{center}
\caption{Shared input microcircuit, in which two neurons receive input
currents with a common component that represents correlated activity or
shared afferents upstream.  Each neuron is a single-compartment
Connor-Stevens model (see Methods), with a maximal A-current conductance
$g_A$ that we vary, eliciting a full range of Type I to Type II spiking
dynamics.  Shared input currents lead to correlated spikes, which are
quantified as shown via spike counts $n_1$, $n_2$ over sliding time windows
of length $T.$  The input currents received by each cell have mean $\mu$ and
fluctuate with total variance $\sigma^2$; the common noise is chosen with
variance $c \sigma^2$, and independent noise terms with variance
$(1-c)\sigma^2$.}
 \label{fig:schematic}
 \end{figure}
%%%%%%%%%%%%%%%%%%%%%%%%%%%%%%%%%%%%%%%%%%%%%%

Neurons throughout the nervous system --- from the
retina~\cite{ShlensRC08},
%{mastronarde,Mei+95,Sch+06,Shl+06}, 
thalamus~(e.g.,~\cite{alonso}),
%bruno,
and cortex~(e.g.,~\cite{zohary94})
%,romo03,Koh+05,deC+96} 
to motoneurons~\cite{Bin+01} --- show temporal correlation between the discharge times of their spikes.
This correlated spiking can impact sensory
discrimination
%~\cite{zohary94,josicSRD09,johnson80,abbott99,averbeck06,Ser+04,kohn07,sompolinsky01},
\cite{averbeck06}
and signal propagation~\cite{Sal+00}.
% and motor behavior~\cite{vaadia}.
  
How do these correlations arise?  We study a simple mechanism in which the inputs to a pair or population of neurons has a common component that is shared across multiple cells (Figure \ref{fig:schematic}).
On an anatomical level, the large number of
divergent connections that span layers and areas makes shared afferents to
pairs of nearby cells unavoidable~\cite{Sha+98}.  Correlated spiking
in areas upstream from the target cells can add to this anatomical factor.  
% cf.~\cite{Kumar:2010p461}).  
In fact, for some neural circuits, shared inputs are
themselves the dominant source of correlated spiking~\cite{Trong2008}.  In
general, correlating effects of shared input interact with effects of recurrent coupling (cf.~\cite{ostojic09}). 
% Taken together, an understanding of the correlating effects shared inputs are fundamental to the generation and propagation of correlated spiking.  

What makes shared input circuitry especially interesting
%for neurophysiology 
is the pivotal role of spike generating dynamics.  
For a given fraction of shared input, these dynamics
control the fraction of shared output --- that is, the fraction of spikes that will be shared across the two cells.
This {\it correlation transfer}
% --- the relationship between input and output (spike) correlations --- 
depends on two factors.  The first is the mechanism of spike generation.  The second is the operating point of the neurons (i.e., their rate and variability of
firing~\cite{Bin+01,RochaDoironSJR07}).  
Excepting~\citetext{hong},
studies of correlation transfer have mostly focused on simplified neuron models, such as integrate-and-fire, phase,
or threshold crossing systems, leaving open allied questions for models with more complex subthreshold and after-spike dynamics.

Here, we study correlation transfer for a family of conductance-based neuron
models that varies from Type I excitability, in which firing can occur at arbitrarily low rates in response to a DC current (as for cortical pyramidal cells), to Type
II excitability, in which firing occurs at a nonzero ``onset" rate (as for
fast-spiking interneurons or the Hodgkin-Huxley model)~\cite{excit,hodgkin48}.  
We use
the Connor-Stevens model~\cite{connor71}, which transitions between Type I and Type
II as $g_A$ -- the maximal conductance of the A-type potassium current -- is
varied (see Fig.~\ref{fig:fIcurves}). 
Beyond firing rates, Type I vs. Type II neurons differ in single-cell computation~\cite{eguPRL07} and synchronization under reciprocal coupling~\cite{excit}.

% -- here, ask complementary 
%questions about differences in correlation driven by common inputs. 

We test the hypothesis --- based on predictions from simplified ``normal form" phase models~\cite{BarreiroST10,marella08,galan} --- that the Type I to Type II transition will produce strong differences in levels and time scales of correlation transfer.   Upon finding a positive result, and explaining it via the filtering properties of individual neurons, we ask how the distinct features
of correlated spiking in Type I vs. Type II neurons manifest in signal transmission through feedforward neural circuits.
Preliminary versions of some findings have appeared in abstract form~\cite{BarreiroCOSYNE09,Shea-BrownSFN09}.

\section*{METHODS}

\subsection*{Circuit setup} We primarily consider the feedforward circuit of
Figure~\ref{fig:schematic}.  Here, each of two neurons receives two sources of
fluctuating current: a common, or ``shared" source $I_c=\sigma \sqrt{c} \, \xi_c(t)$, and a
``private" source $I_1=\sigma \sqrt{1-c} \,  \xi_1(t)$ or $I_2=\sigma
\sqrt{1-c} \, \xi_2(t)$. Each of these inputs is chosen to be a scaled statistically
independent, Gaussian white noise process (uncorrelated in time) --- that is, $\langle \xi_i(t) \xi_i(t+\tau) \rangle = \delta(\tau)$ for $i=1,2,c$.  This
is for simplicity and agreement with prior studies of correlated
spiking~\cite{Lin+05,RochaDoironSJR07,SBJRD07,marella08,Vilela:2009p370,BarreiroST10}.  The common current $I_c$ has variance $\sigma^2
c $; each private current has zero mean and variance $\sigma^2 (1-c)$.
Note that these scalings are chosen so that the total variance of current
injected into each cell is always $\sigma^2$, while the parameter $c$ gives the
fraction of this variance that arises from common input sources.  For example,
when $c=0.5$, 50\% of each neuron's presynaptic inputs come from the shared
and 50\% from the independent input.  Finally, the mean of the total current
received by each cell is given by $\mu$.  This term represents the total bias
toward negative or positive currents from all sources; in
Fig.~\ref{fig:schematic}, it is  illustrated as part of the common
input for simplicity.

The combined currents, \begin{eqnarray*} I_{app,i}(t) &=&\mu + I_c(t) + I_i(t) \\ &= &\mu +  \sigma \sqrt{c}\,  \xi_c(t)  + \sigma \sqrt{1-c}
\, \xi_i(t)  \label{eqn:Iapp} \end{eqnarray*}
($i=1,2$) are injected into identical single-compartment,
conductance-based membrane models (see Methods, ``Neuron model"); spike
times are identified from the resulting voltage trace.

 \subsection*{Neuron model}  \label{sec:model} 
 
 We
investigate correlation transfer in the Connor-Stevens model, which was
designed to capture the low firing rates of a crab motor axon \cite{connor71,connor77}. This model adds a transient potassium
current, or $A$-current, to sodium and delayed-rectifier potassium currents of Hodgkin-Huxley type.  The A-Type channel
provides extended after-spike hyperpolarization currents, which lead to arbitrarily low firing rates and hence Type I excitability (see Introduction and  Methods, ``Characterizing the dynamics of spike generation," below).

The voltage equation is
\begin{eqnarray}
 C_{M} {\frac{dV}{dt}} &=&-g_{L}(V-E_{L})-g_{Na}m^{3}h(V-E_{Na})-g_{K}n^{4}(V-E_{K})  \label{eqn:CS_voltage}\\
 & - & g_A A^3 B (V-E_K) + I_{app},   \nonumber
\end{eqnarray}
The gating variables $m$, $n$, $h$, $A$, and $B$ each evolve according to the
standard voltage-gated kinetics; e.g., for $m$: 
\begin{eqnarray}
\frac{dm}{dt} & = & \frac{m_{\infty}(V) - m}{\tau_m(V)} \label{eqn:CS_gating}
\end{eqnarray} 
where $m_{\infty}(V)$ is the steady-state value and $\tau_m(V)$ is the (voltage-dependent) time constant.  All equations and parameters are exactly as specified as in
\citetext{connor77}, with the exception that we vary the maximal $A$-current 
conductance, $g_A$, over the range of values reported below.
As $g_A$ is decreased from the value set by 
\citetext{connor77}, the neuron transitions from Type I to Type II excitability; 
we describe this phenomenon next.

%%%%%%%%%%%%%
 \begin{figure}[t!]
 \vspace{-.25cm}
     \begin{center}
 {\includegraphics[width=6in]{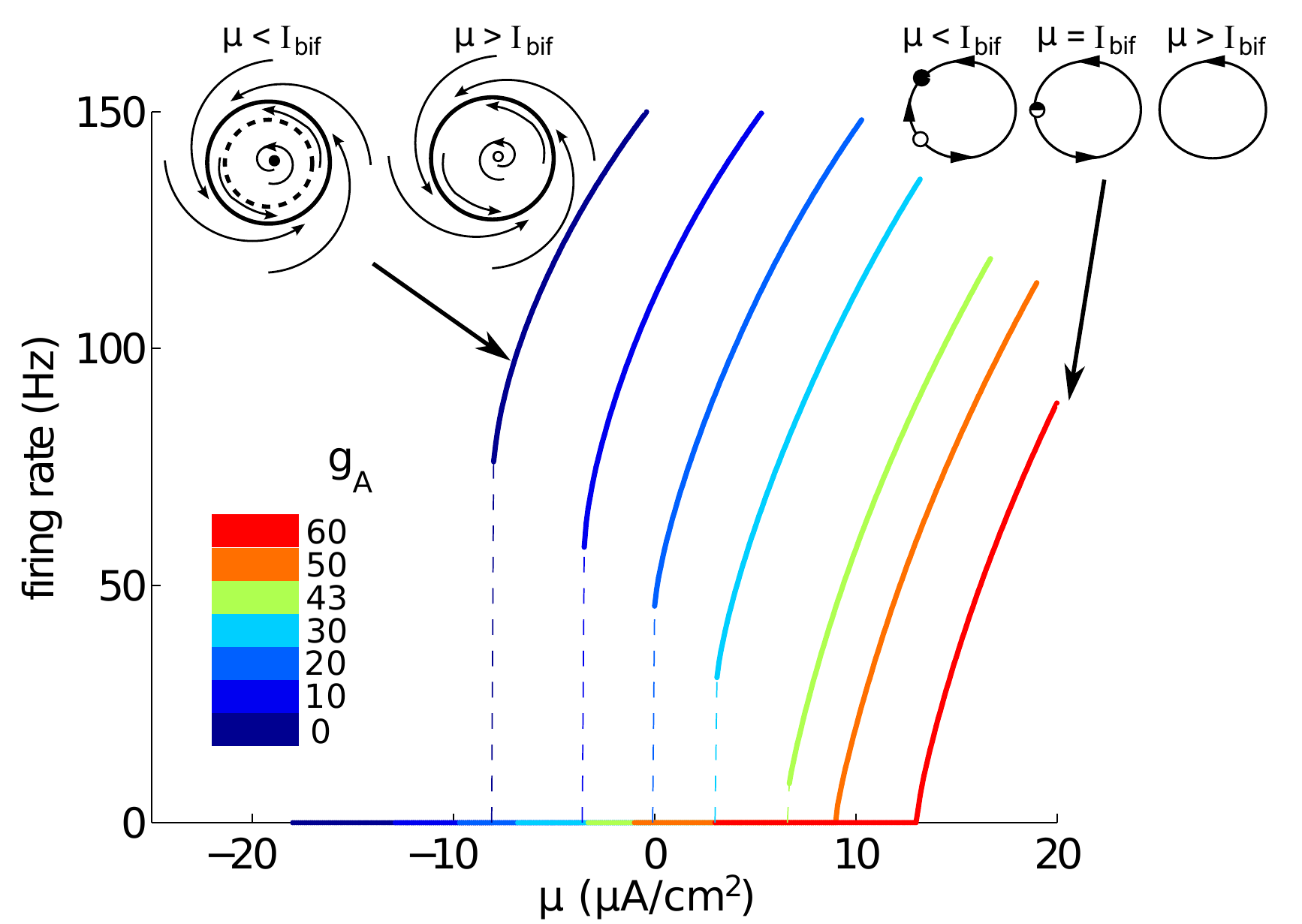}}
     \end{center}
\caption{Firing rate vs. injected current ($f-I$) curves, for the deterministic ($\sig =0$) Connor-Stevens model. Several values of $g_A$, yielding a range from Type II to Type I excitability, are shown --- note the nonzero ``onset" firing rates and Type II excitability for $g_A \approx 0$ ${\rm mS}/{\rm cm}^2$, zero onset rate and Type I excitability for $g_A \approx 60$ ${\rm mS}/{\rm cm}^2$, and a gradual transition between. 
Insets show cartoons of dynamical transitions that lead to non-zero vs. zero onset rates: a subcritical Hopf bifurcation (left) and a saddle-node on invariant circle bifurcation (right).
}
 \label{fig:fIcurves}
 \end{figure}

%%%%%%%%%%%%%%%%%%%%%%%%%%%%%%%%%%%%%%%%%%%%%%

\subsection*{Measuring spike train correlation}

We represent the output spike trains as sequences of impulses $y_i(t) =
\sum_i \delta(t-t_i^k)$, where $t_i^k$ is the time of the $k$th spike of the
$i$th neuron.  The firing rate of the $i$th cell, $\langle y_i(t) \rangle$,
is denoted $\nu_i$. To quantify correlation over a given
time scale $T$, we compute the Pearson's correlation coefficient of spike
counts over a time window of length $T$ (as in, e.g.,~\cite{zohary94,bair01}):
\begin{eqnarray*} \rho_T & = &
\frac{\Cov(n_1,n_2)}{\sqrt{\Var(n_1)}\sqrt{\Var(n_2)}} \end{eqnarray*} where
$n_1$, $n_2$ are the numbers of spikes simultaneously output by neurons 1 and 2 respectively in a time window of length $T$; i.e. $n_i(t) = \int_t^{t+T} y_i(s) \, ds$. If
$\rho_{T}$ is measured at values of $T$ that are less than a typical inter-spike interval
(ISI), we are essentially measuring the degree of synchrony between
individual spikes.  For larger $T$ values, $\rho_{T}$ assesses total
correlation between numbers of spikes emitted by each cell.

A short calculation (cf.~\cite{bair01,Cox+66}) shows that $\Cov(n_1,n_2)$ is
\begin{eqnarray} 
\Cov(n_1,n_2) & = & T \int_{-T}^T C_{12}(t) \frac{T - |t|}{T} \, dt \label{eqn:Cov_from_C} 
\end{eqnarray} 
where the spike train
cross-covariance $C_{12} (\tau) = \langle y_1(t) y_2(t + \tau) \rangle -
\nu_1 \nu_2$.
Similarly, the variance $\Var(n_1)$ can be given in terms of the spike train
autocovariance function.  The autocovariance function of neuron 1, defined as
$A_{1} (\tau) = \langle y_1(t) y_1(t+ \tau) \rangle - \nu_1^2$, satisfies
\begin{eqnarray*} 
\Var(n_1) & = & T \int_{-T}^T
A_{1}(t) \frac{T - |t|}{T} \, dt \; ,  
\label{eqn:Var_from_C} \end{eqnarray*}
and similarly for neuron 2.

\subsection*{Characterizing the dynamics of spike generation}

We give a brief discussion of spike-generation dynamics in the Connor-Stevens
system.  We organize this by focusing on how the neurons transition from quiescent (i.e., rest) behavior to periodic spiking in response to constant, noiseless (DC) current (i.e., when fluctuation terms $\sigma = 0$).  While these results are well-established~\cite{rush95} we review critical aspects that will help us understand how correlated spiking arises in the circuit of Fig.~\ref{fig:schematic}.

For any value of $g_A$, there is a critical value of the DC current, $\mu=I_{bif}(g_A)$, such that the neuron has a stable stationary state --- that is, it remains at a resting voltage --- if
$\mu <  I_{bif} (g_A)$. If $\mu > I_{bif}(g_A)$, then the neuron instead spikes periodically. We refer to these two regimes as
\textit{subthreshold} and \textit{superthreshold} respectively. If the
current $\mu$ is ramped slowly from a subthreshold value, the membrane
potential will shift slowly until $\mu = I_{bif}(g_A)$, at which point periodic spikes begin; the rate continues to increase with $\mu$ over some range, as quantified by 
familiar firing rate-current ($f-I$) curves (Fig.~\ref{fig:fIcurves}).

As discussed above, neurons are often classified as Type I vs. Type II based on whether their 
%firing rate-current 
$f-I$
curves are continuous vs. discontinuous (with a jump) at $\mu = I_{bif}$.  Figure~\ref{fig:fIcurves} shows that the Connor-Stevens model is Type II for  $g_A \approx 0$ ${\rm mS}/{\rm cm}^2$, Type I for $g_A \approx 60$ ${\rm mS}/{\rm cm}^2$, and displays a gradual transition in between.
This qualitative shift in behavior is 
related to the underlying model~(Equations \ref{eqn:CS_voltage},\ref{eqn:CS_gating})
through \textit{bifurcation theory}~\cite{izhikevich07,excit}. The key fact is that all possible changes from resting to periodic spiking behavior 
%, the equations undergo a qualitative change in their behavior, which may be characterized by analyzing the equations at the resting point.
fall into a few qualitatively equivalent categories of bifurcation.  We next describe two of these categories that we will put to use below.

In a \textit{saddle-node on invariant circle} (SNIC) bifurcation, the stable resting state merges with an \textit{unstable} resting state to create a single steady state precisely when 
$\mu = I_{bif}(g_A)$.  A solution to Equations (\ref{eqn:CS_voltage},\ref{eqn:CS_gating}) that begins near this steady state can return to it, via an excursion through state space that traces out a spike.  This spiking trajectory persists for $\mu > I_{bif}$.  However, due to the steady states at nearby values of $\mu$, spiking trajectories take a long time to return to where they started --- technically an infinite
amount of time when $\mu = I_{bif}(g_A)$; see the righthand schematic in Figure~\ref{fig:fIcurves}.  As a result, the firing rate for $\mu$ near $I_{bif}(g_A)$ is arbitrarily low, resulting in a continuous $f-I$ curve and Type I excitability.  

In a \textit{subcritical Hopf} bifurcation, by contrast, the stable resting state instead loses stability as an unstable periodic orbit 
shrinks into this point. For neural systems, there is typically also a stable periodic orbit in the state space.
%, which the system has not been able to access because of the stable (attracting) resting state.  
Once the resting state becomes unstable, trajectories quickly diverge from the resting state to the stable periodic orbit, which has a non-zero frequency $f_{bif}$; see the lefthand schematic in Figure~\ref{fig:fIcurves}.  This results in a discontinuous $f-I$ curve that jumps from zero (at rest) to $f_{bif}$ at $I_{bif}(g_A)$ --- the characteristic of Type II excitability.

We use software tools to automate the bifurcation analysis of the Connor-Stevens model: specifically XPP \cite{XPPbook02}, and MatCont \cite{Matcont03}. This allows us
to track fixed points and limit cycles as system parameters $g_A$, $\mu$ vary,
and to check the mathematical conditions that define bifurcation types~\cite{GH}.
\bigskip

\subsection*{Linking linear response theory, spike-triggered averages, and spike count correlations}

When the variance $c$ of the shared input is small (see Fig.~\ref{fig:schematic}),
then we can treat the circuit with a shared input as a perturbation from two independently firing neurons. We describe this
perturbation via {linear response theory} \cite{Lin+05,RochaDoironSJR07,ostojic09,Ostojic:2011p423}, which is related to classical Linear-Poisson (LP) models of neural spiking ~\cite{PerkelGM67}.  That is, we make the assumption that the {\it change} in a neuron's instantaneous firing rate $\nu_i(t)$ due to the shared input signal can be represented
by linearly filtering the common (perturbing) input $I_c$:
\begin{eqnarray}
\nu_i(t) & \equiv & \langle y_i(t) | I_c \rangle \nonumber \\
& = & \nu_{0,i} + \int_0^{\infty} K(s) I_c(t-s) \, ds  \nonumber \\
& = & \nu_{0,i} + (K \ast I_c) (t) 
\label{eqn:filter1}
\end{eqnarray}
where the filter $K(t) = 0$ for $t < 0$ (causality) and $\nu_{i,0}$ is the ``background" average firing rate of the independently firing neuron.

Eqn.~\eqref{eqn:filter1} is extremely useful, because it isolates the common component of the response of neurons $i=1$ and $i=2$, which is an enhanced (or depressed) tendency to emit spikes, at a rate determined by the filtered, common input.  As a result, the cross-covariance function, 
\begin{eqnarray}
C_{12}(\tau) & = & \langle (\nu_1(t) - \nu_{0,1}) (\nu_2(t+\tau) - \nu_{0,2}) \rangle  \nonumber
\end{eqnarray}
becomes
\begin{eqnarray*}
C_{12}(\tau) & = & \int_0^{\infty} \int_{-s}^{\infty} K(s) K(s+r) \langle I_c(t-s) I_c(t-s+\tau-r) \rangle \, dr \, ds \nonumber
\end{eqnarray*}
This simplifies:  because we assume $I_c$ is white (uncorrelated in time), $\langle I_c(t) I_c(t+\tau) \rangle = \Var(I_c) \delta(\tau)$.  $\Var(I_c)$ is the instantaneous variance of the noisy process (here assumed to be constant), so that
\begin{eqnarray}
C_{12}(\tau) & = & \Var(I_c)  \int_0^{\infty} K(s) K(s+\tau) \, ds \nonumber   \\
& = & \Var(I_c) (K \ast \tilde{K}) (\tau)  \label{eqn:C_from_K}
\end{eqnarray}
where $\tilde{K}(t) \equiv K(-t)$; cf.~\cite{ostojic09}.

Thus, the cross-covariance between our two spike trains has a simple expression in terms of the filter with which neurons process the common input into a firing rate.  It remains only to identify this filter $K(t)$.  As in classical cases, this is precisely given by a {\it spike triggered average}~\cite{GK98}.  Different from the classical setting, but as in~\citetext{ostojic09}, it is only the common component of the input current that is averaged in this procedure.

The connection to the spike-triggered average may be seen by looking at the response to a white noise stimulus.
On the one hand,
\begin{eqnarray}
\langle I_c(t) \nu_i(t+\tau) \rangle & = &  \langle I_c(t) (K \ast I_c)(t+\tau) \rangle  \nonumber \\
& = & \left( K \ast \langle I_c(t) I_c(t+\tau) \rangle \right) (\tau) \nonumber \\
& = & \Var (I_c) K(\tau) \nonumber
\end{eqnarray}
if the noise is white. On the other, 
by ergodicity, averaging over repeated presentations {\it and} different stimuli $I_c$ will yield
a response equivalent to averaging over repeated presentations of a single long (duration $T_{max}$) stimulus.  As $T_{max} \rightarrow \infty$,
\begin{eqnarray}
\langle I_c(t) \nu_i(t+\tau) \rangle & = &  \frac{1}{T_{max}} \int_0^{T_{max}} I_c(t) \langle y(t+\tau) | I_c \rangle \, dt \nonumber \\
& = &  \frac{1}{T_{max}} \int_0^{T_{max}} I_c(t) \left \langle  \sum_{k=1}^N \delta(t+\tau-t_k)  \right \rangle \, dt \nonumber\\
& = & \frac{N}{T_{max}} \left( \frac{1}{N}\sum_{k=1}^N I_c(t_k-\tau) \right) \nonumber\\
& = & \nu_i \left( \frac{1}{N}\sum_{k=1}^N I_c(t_k-\tau) \right)
\end{eqnarray}
which is exactly the average stimulus preceding a spike (multiplied by the firing rate).
Therefore the linear response function $K(\tau)$ is related to the spike-triggered average $\STA(\tau)$ as follows:
\begin{eqnarray}
K(\tau) & = &  \frac{\nu_i}{\Var(I_c)} \left( \frac{1}{N}\sum_{k=1}^N I_c(t_k-\tau) \right) \nonumber \\ & = &   \frac{\nu_i}{\Var(I_c)} \STA(\tau) \, .   \label{eqn:K_from_STA}
\end{eqnarray}
Below, we use this expression to derive $K(\tau)$ from numerically computed spike triggered averages.

\subsection*{Relating spike-triggered averages and spike-generating dynamics}
In order to relate the common input STA (defined in the previous subsection) to spike generating dynamics,
it will be helpful for us to derive an explicit formula for the common input STA of a \textit{phase model}, 
which captures the response of a tonically spiking neuron to a small-amplitude current $I(t)$.
Our formula, and the calculation that yields it, is very similar to a relationship previously derived \cite{eguPRL07} for the STA of a 
phase oscillator {\it without} background noise.

We consider a model which tracks only the phase of a neuron as it progresses along its periodic spiking orbit:
\begin{eqnarray}
\frac{d \theta}{dt} & = & \omega + Z(\theta) I(t), \qquad \theta \in [0,2\pi)
\end{eqnarray}
The function $Z(\theta)$, called a \textit{phase response curve} or 
PRC \cite{ermentrout84,geomtime,ErmentroutTBook,ReyesF93}, determines how a brief current injection
applied at a specific phase of the cycle affects the timing of the next spike. By convention, the neuron is said to ``spike"
when $\theta$ crosses $2 \pi$.

To begin, we assume that the phase model is forced by scaled zero-mean, stationary stochastic
processes, which we also label $\xi_c(t)$ and $\xi_i(t)$. For now, $\xi_c(t)$ and $\xi_i(t)$ have unit variance and are differentiable with some finite correlation time $\tau$,
although we will consider the limit  $\tau \rightarrow 0$ (i.e. the white noise limit).
We are interested in the average value of $\xi_c(t)$ that precedes a spike;
the term $\xi_i(t)$ will play the role of a background noise.
Assuming that the background noise process is scaled by a small constant $\sig$, and that $\xi_c$ is scaled by an order of magnitude $\epsilon$ smaller still, we
write the evolution equation of the phase model as
\begin{eqnarray}
\frac{d\theta}{dt} & = & 1 + Z(\theta) (\sigma \xi_i(t) + \sigma \epsilon \xi_c(t)), \qquad \theta \in [0,T) \nonumber
\end{eqnarray}
where $\sigma, \epsilon \ll 1$.
Note that we have chosen our phase variable to have unit speed; i.e $\theta \in [0,T)$, where $T$ is the period of the unperturbed ($\sigma = 0$) oscillator.
We proceed as in \citetext{eguPRL07}: writing $\theta$ as a series in the small parameters $\sigma$ and $\epsilon$
\begin{eqnarray}
\theta(t) & = & \theta_0(t) + \sigma \theta_{10}(t) + \epsilon \theta_{01}(t) + \sigma^2 \theta_{20}(t) + \sigma \epsilon \theta_{11}(t) + \epsilon^2 \theta_{02}(t)+ ... \nonumber
\end{eqnarray}
and matching terms of same order in the evolution equation, we find $\theta_0(t) = t$. We additionally find that
$\theta_{01}=0$, $\theta_{02}=0$, and
\begin{eqnarray*}
\theta_{10}'& = & Z(t)\xi_i(t)\\
\theta_{20}'&=& Z'(t)\xi_i (t) \theta_{10}(t)\\
\theta_{11}'&=& Z(t) \xi_c(t)
\end{eqnarray*}
so that
\begin{eqnarray*}
\theta_{10}(t) & = & \int_0^t Z(s) \xi_i(s) \, ds\\
\theta_{20}(t) & = & \int_0^t Z'(s) \xi_i(s) \int_0^s Z(r) \xi_i(r) \, dr \, ds\\
\theta_{11}(t) & = & \int_0^t Z(s) \xi_c(s) \, ds \;.
\end{eqnarray*}
In order to compute the spike-triggered average, we need to find the time of the next spike, assuming the neuron has just spiked ($\theta(0)=0$); in other words, the time $\tau$ when $\theta(\tau)=T$. As above, we expand
\begin{eqnarray}
\tau & = & T + \sigma \tau_{10} + \sigma^2 \tau_{20} + \sigma \epsilon \tau_{11} + ...
\end{eqnarray}
Using our previous expressions for $\theta(\tau)$, and using the fact that $\tau = T + \sigma \tau_{10} + O(\sigma^2, \sigma \epsilon)$ to
decompose the stochastic integrals, we find
\begin{eqnarray*}
\tau_{10} & = & -\int_0^T Z(s) \xi_i(s) \, ds\\
\tau_{20} & = & -\int_{T}^{T+\sigma \tau_{10}} Z(s) \xi_i(s) \, ds - \int_0^T Z'(s) \xi_i(s) \int_0^s Z(r) \xi_i(r) \, dr \, ds\\
\tau_{11} & = & -\int_0^T Z(s) \xi_c(s) \, ds \;.
\end{eqnarray*}
Next, we use Taylor's theorem for smooth functions to expand $\xi_c$ about $T-t$ to compute
\begin{eqnarray*}
\STA(t) & = & \langle I_c(\tau - t) \rangle \\
& = & \sigma \epsilon \langle \xi_c(T +\sigma \tau_{10} + \sigma^2 \tau_{20} + \sigma \epsilon \tau_{11} - t \rangle\\
& = & \sigma \epsilon \langle \xi_c(T-t) + (\sigma \tau_{10} + \sigma^2 \tau_{20} + \sigma \epsilon \tau_{11}) \xi'_c(T -t) \\
&& +  \frac{1}{2} (\sigma \tau_{10} + \sigma^2 \tau_{20} + \sigma \epsilon \tau_{11})^2 \xi''_c(T -t)  \rangle\\
& = & \sigma \epsilon \langle  \xi_c(T-t) + (\sigma \tau_{10} + \sigma^2 \tau_{20} + \sigma \epsilon \tau_{11}) \xi'_c(T -t) \\
&& +  \frac{1}{2} \sigma^2 \tau_{10}^2  \xi''_c(T -t)  \rangle + O(\sigma^3, \sigma^2 \epsilon).
\end{eqnarray*}
where we have kept terms up to second order both in our expression for $\tau$, and in our Taylor expansion of $\xi_c$.
We can use the independence of $\xi_c$ and $\xi_i$ to eliminate a large number of terms, as 
\begin{eqnarray*}
\langle \xi_c (t) \xi_i(t+s) \rangle & = & \langle \xi_c(t) \rangle \langle \xi_i(t+s) \rangle 
 =  0.
\end{eqnarray*}
Similarly,
\begin{eqnarray}
\langle \xi_c (t) \xi'_i(t+s) \rangle & = & \langle \xi_c(t) \rangle \langle \xi'_i(t+s) \rangle = 0,
\end{eqnarray}
and so forth for expressions with higher derivatives. The only term that survives is
\begin{eqnarray*}
\STA(t) & = & \sigma \epsilon \langle \xi'_c(T-t) \times - \sigma \epsilon \int_0^T Z(s) \xi_c (s) \, ds \rangle\\
& = & -(\sigma \epsilon)^2 \int_0^T Z(s) \langle \xi'_c (T-t) \xi_c (s) \rangle \, ds\\
& = &  -(\sigma \epsilon)^2 \int_0^T Z(s) A_c'(T-t-s)  \, ds\\
& = &  -(\sigma \epsilon)^2 \int_0^T Z'(s) A_c(T-t-s) \, ds,
\end{eqnarray*}
where $A_c$ is the autocovariance function of $\xi_c$ and we used integration by parts in the final step.
Taking the white noise limit ($A_c(T-t-s) \rightarrow \delta(T-t-s)$) and using the periodicity of the PRC ($Z(T-t) = Z(-t)$),
we recover a very similar expression as \citetext{eguPRL07}:
\begin{eqnarray}
\STA(t) & = & -(\sigma \epsilon)^2 Z'(-t). \label{eqn:STAPRC}
\end{eqnarray}

\subsection*{Numerical simulations and estimates of spike count statistics} To compute spike count correlations and
other statistical quantities, we performed Monte Carlo simulations of the circuit  
in Fig.~\ref{fig:schematic}. The governing Connor-Stevens equations (\ref{eqn:CS_voltage},
\ref{eqn:CS_gating}) were integrated using the stochastic Euler method with time step
$\Delta t = 0.01$ ms, for a total time $T_{max}$ of $8 \times 10^6$ ms.
Random input currents were chosen at
each time step using a standard random number generator~\cite{mz94}. To
facilitate exploration of parameter space $\mu$, $\sigma$, $c$, and $g_A$, we
distributed computations on parallel machines through the NSF
Teragrid program (http://www.teragrid.org). The simulation code was
implemented in FORTRAN90, and distribution scripts in Python for running on clusters with and without PBS submission protocols.  All code and scripts will be available at the modelDB site upon publication (http://senselab.med.yale.edu/modeldb/).

We register spikes in our simulations at times when the membrane voltage exceeds $- 30$ mV and maintains a positive slope in voltage for the next three time steps (0.03 ms). To avoid counting each spike more than once, we omit a 2 ms refractory period after each spike.
%The voltage threshold is set high enough to avoid the biggest subthreshold fluctuations.

Spike count statistics were computed directly from the recorded spike times, based on 
%We assume that the spike count $n_j$ over a fixed time window $T$ is stationary; therefore we use spike counts over 
a single long simulation, after discarding an initial transient (200 ms).
When sampling spike counts over a time window $T$, we advance the window by $\frac{1}{4}T$,
resulting in approximately $4 T_{max}/T$ (correlated) samples; consequently, our estimates of spike counts become noisier as $T$ increases.
To estimate standard errors on spike count statistics, we further divided the simulation into 10 equal time intervals ($8 \times 10^5$ ms each)
and computed statistics on each sub-simulation; the standard deviation, divided by $\sqrt{10}$, gives us an estimated standard error of the mean.
When appropriate, these are presented along with the mean estimates, as error bars.

Below, we also report spike triggered averages (STAs) described above; these were computed using long simulations of length $8 \times 10^7$ ms for several sets of parameter values $\mu$, $\sigma$, and $g_A$.  To compute these, the common input current $I_c$ was treated as the ``signal" which was averaged and the private input as a ``background" which was not.  We used $c=0.10$ in this computation.  In our code, the history $I_c$ 
was continuously recorded for a duration into the past; when a spike was recorded, the STA was augmented by this current.

\begin{figure}[tp!]
 \vspace{-.25cm}
     \begin{center}
 {\includegraphics [width=1\textwidth]{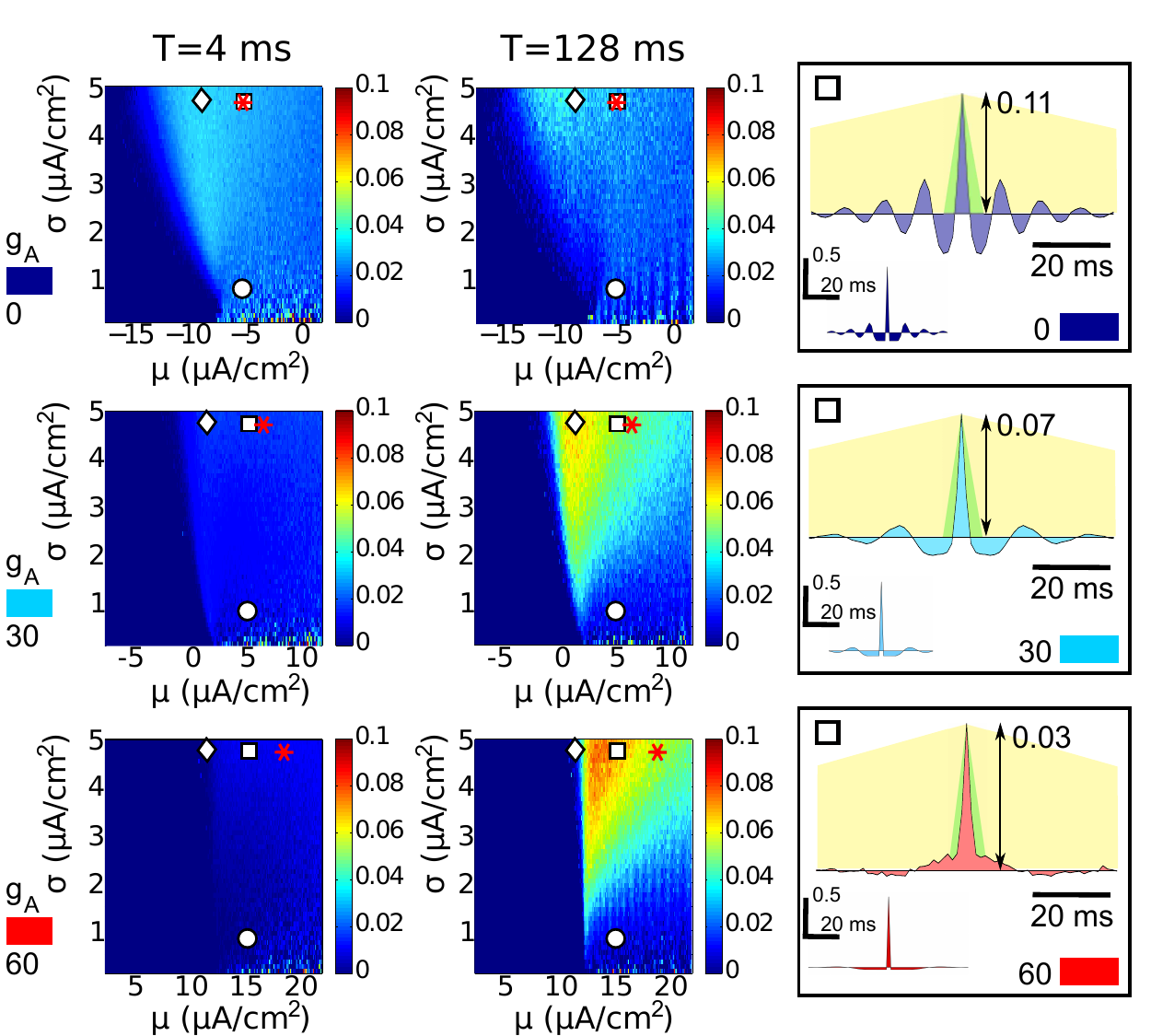}} 
\caption{Spike count correlations for three models at both short and long time scales. Each row displays data from a value of $g_A$: from top to bottom, Type II ($g_A=0$ ${\rm mS}/{\rm cm}^2$), 
intermediate ($g_A=30$ ${\rm mS}/{\rm cm}^2$), and Type I ($g_A=60$ ${\rm mS}/{\rm cm}^2$).  Left column:  Spike count correlations $\rho_T$, for short windows $T=4$ ms.  Center:  Spike count correlations $\rho_T$, for long windows $T=128$ ms.  
Markers indicate points used for cross-model comparison: subthreshold by 1 ${\rm \mu A}/{\rm cm^2}$ (diamonds), superthreshold by 2 ${\rm \mu A}/{\rm cm^2}$ and low noise (circles), superthreshold by 2 ${\rm \mu A}/{\rm cm^2}$ and high noise (squares), and superthreshold with matched Fano factor (stars). Right: Cross-covariance and autocovariance (inset) functions for the superthreshold high noise points (squares). Behind cross-covariance functions, the shape of the triangular kernel that relates this function to spike count covariance (as in Eqn. \ref{eqn:Cov_from_C}) is illustrated for $T=4$ ms (green) and $T=128$ ms (yellow). For each value of $g_A$, autocovariance functions are given in normalized units (so that $A(0) = 1$).  }\label{fig:corrpanels}
\end{center}
 \end{figure}

Finally, we generate auto- and cross-correlograms (shown in Figure \ref{fig:corrpanels}) by collecting inter-spike intervals (ISIs) from our simulations in 1 ms-long bins.
These are used, after the standard normalization, as auto- and cross-covariance functions.
%
%\bei
%	\item Euler method, timestep
%	\item Code written in fortran, freely available on modelDB (we will post it at time of paper submission, or at least post it on our group webpage
%	\item Teragrid
%	\item Errorbars, number of `trials,' runlengths
%	\item STA calcs
%\eei

\section*{RESULTS} %%%%%%

%%%%%%%%%
\begin{figure}[t!]
  \vspace{-.25cm}
     \begin{center}
     \includegraphics{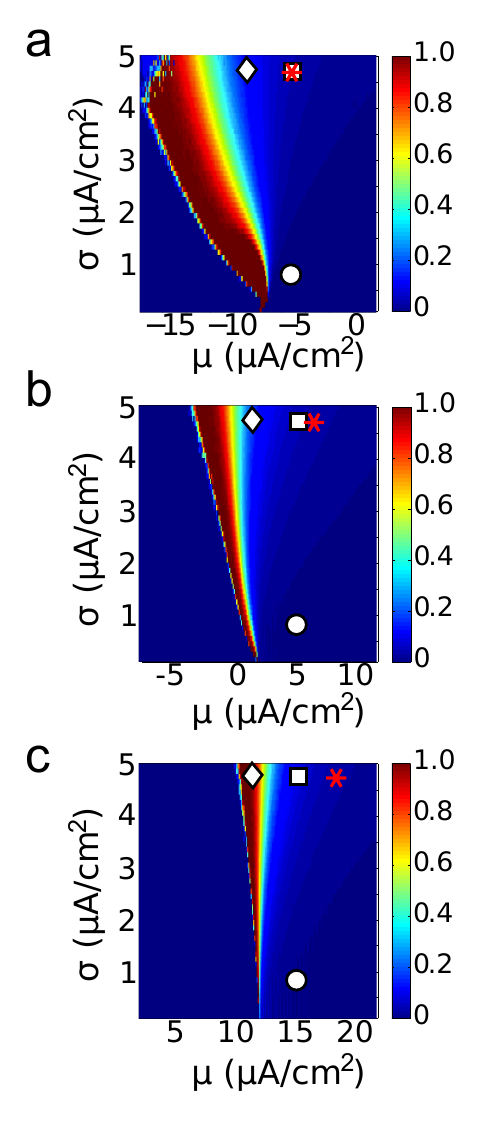}
     \end{center}
 \vspace{-.80cm}
 \caption{Fano factor of spike counts over a long time window ($T=256$ ms) for a $\sim200 \times 50$ grid of values for the mean current $\mu$ and variance $\sigma$. 
From top to bottom, Type II to Type I: (a) $gA=0$,
(b) $gA = 30$, and (c) $gA=60$ ${\rm mS}/{\rm cm}^2$. 
Markers indicate relative location of $(\mu, \sigma)$-pair; subthreshold by 1 ${\rm \mu A}/{\rm cm^2}$(diamond), superthreshold by 2 ${\rm \mu A}/{\rm cm^2}$ with low noise (circle) and high noise (square), and superthreshold with matched Fano factors (asterisk, see text).}
 \label{fig:Fanofactor}
 \end{figure}
%%%%%%%%%%%%%%%%%%%%%%

%%%%%%%%%%%%%%%%%%%%%%%%%%%%

%%%%%%%
\subsection*{Rich structure of spike count correlations over short and long time scales}

Our central findings contrast how different conductance-based neuron models produce correlated spiking when they receive overlapping fluctuating inputs, via the shared-input circuitry in Fig.~\ref{fig:schematic}.  Specifically, we show how this correlation depends on the Type I vs. Type II excitability class of a neuron described by the well-studied Connor-Stevens model.  As discussed above, neurons are often classified as Type I vs. Type II based on whether their firing rate-current curves are continuous vs. discontinuous at $\mu = I_{bif}$.  Figure~\ref{fig:fIcurves} demonstrates --- as shown in~\cite{rush95} --- that the Connor-Stevens model is Type II when the maximal A-current conductance $g_A \approx 0$ ${\rm mS}/{\rm cm}^2$, Type I for $g_A \approx 60$ ${\rm mS}/{\rm cm}^2$, and displays a gradual transition in between.  Thus, we fix the neurons in the shared-input circuit to a point along the spectrum from Type I to Type II excitability by choosing different values of $g_A$. 

%on the strengths $\mu$ and $\sigma$ of the DC and fluctuating components of the input currents

To compute levels of correlated spiking, we then fix the correlation in the input currents --- that is, the fraction of the current variance that is shared vs. private to the two cells --- to a preset value $c$.  For each value of $c$ and $g_A$, we compute spike count correlations for wide range of operating points for the neurons, as determined by a $\sim200 \times 50$ grid of values for the mean current $\mu$ and variance $\sigma$ (both $\mu$ and $\sigma$ are sampled at a resolution of $0.1$ ${\rm \mu A}/{\rm cm^2}$).  
Specifically, we vary $\mu$ over values centered at the threshold current $I_{bif}(gA)$, from a minimum $\mu=I_{bif}(gA) - 10$ $({\rm \mu A}/{\rm cm^2})$ to a maximum $\mu =  I_{bif}(gA) + 10$ $({\rm \mu A}/{\rm cm^2})$ for each value of $g_A$.  This enables us to cover, respectively, both subthreshold (i.e., fluctuation-driven, $\mu<I_{bif}(gA)$) and superthreshold (i.e., mean-driven, $\mu>I_{bif}(gA)$) firing regimes for each value of $g_A$.  We additionally vary $\sigma$ over  $0 < \sigma \leq 5$ $({\rm \mu A}/{\rm cm^2})$, so that we cover the range from nearly Poisson, irregular spiking to nearly periodic, oscillatory spiking.  This is demonstrated by Figure \ref{fig:Fanofactor}, which shows the Fano factor of spike counts over a long time window ($T=256$ ms) --- a proxy for the squared inter-spike interval coefficient of variation~\cite{GK98} ---  over the entire $\mu$, $\sigma$ parameter space for three representative values of $g_A$ (0, 30, 60  ${\rm mS}/{\rm cm}^2$). 
For each value of $g_A$, the Fano factor spans a range from near zero (periodic) to one (i.e. Poisson-like) or higher.

For each set of parameters $g_A$, $c$, $\mu$ and $\sigma$, we compute the Pearson's correlation coefficient $\rho_T$ between the spike counts that the neuron pair in Fig.~\ref{fig:schematic} produces in time windows of length $T$.  Figure \ref{fig:corrpanels} summarizes the results, for inputs with 10\% shared variance ($c=0.1$).  Here, we view $\rho_T$ over the entire $\mu$, $\sigma$ parameter space for three representative values of $g_A$ (0, 30, 60 ${\rm mS}/{\rm cm}^2$) and two different time windows ($T = 4$ ms and $T=128$ ms).  Values of $\rho_T$ depend in a strong but systematic way on all of the parameters we have introduced. As we move down a column, we see major qualitative differences in levels of correlation that emerge at different points through the Type II ($g_A=0$) to Type I ($g_A=60$) spectrum.  Within each panel, the operating point set by input mean and variance ($\mu, \sig$) have a strong impact on $\rho_T$.   Finally, the levels and trends in $\rho_T$ depend strongly on the time scale $T$.  We now describe these trends in more detail; the sections that follow will give an explanation for how they arise.

We begin with the upper panels in Fig.~\ref{fig:corrpanels}, which show correlation $\rho_T$ for $g_A=0$ and hence Type-II excitability.  First, note that correlations are overall quite weak.  The largest values of $\rho_T$ obtained are $\approx 0.04$, indicating that $\approx 40\%$ or less of correlations in input currents are ever transferred into correlations in output spikes.    Moreover, the level of correlations $\rho_T$ and their dependence on input parameters $\mu$ and $\sigma$ appear roughly similar for both short and long time scales $T$.  In both cases, for a fixed value of DC input $\mu$, a general trend is that $\rho_T$ gradually increases with fluctuation strength $\sigma$.  For a fixed value of $\sig$, in general $\rho_T$ first increases and then decreases with $\mu$; the dependence is slightly more complex at longer $T$.  Significantly non-zero values of $\rho_T$ are present for $\mu < I_{bif}$, as $\sigma$ becomes appreciably high; this reflects the bistable firing dynamics of the underlying deterministic system, which supports both a stable resting state and a stable spiking trajectory for $\mu < I_{bif}$ (see Methods, ``Characterizing the dynamics of spike generation").

For Type I excitability at $g_A=60$ (lower panels in Fig.~\ref{fig:corrpanels}), the picture is dramatically different.  First, there is a marked difference between correlation elicited on short vs. long time scales $T$, with much stronger correlations observed for larger $T$.  Moreover, correlations produced by Type I neurons over longer time scales $T$ are much higher than those observed for Type II neurons at any time scale:  the largest values of $\rho_T$ obtained for Type I are $\approx 0.8$, indicating that $\approx 80\%$ of correlations in input currents can be transferred into spike correlations.  Conversely, correlations for Type I neurons are strongly suppressed at short time scales, where $\lsim 10\%$ of input correlations are transferred.  Overall, trends in $\rho_T$ as $\mu$ and $\sig$ vary are similar to those found previously:  correlations increase with $\sigma$, and first increase --- then decrease --- with $\mu$.  

Correlation transfer in the intermediate model, $g_A=30$, displays trends between those of the Type I ($g_A=60$) and Type II ($g_A=0$) cases.  
%while retaining some features of the Type II ($g_A=0$) case.  
As when $g_A=60$, spike count correlations $\rho_T$ are very low for short time windows $T$ and attain intermediate to high values $\approx 60\%$ of input correlations transferred for longer $T$.  As for both $g_A=60$ and $g_A=0$, $\rho_T$ increases with noise magnitude $\sigma$ and displays a nonmonotonic trend with mean current $\mu$.  
%%%%%%%%%%%%%%%%%%%%%%%%%%%%
 \begin{figure}[ht!]
  \vspace{-.25cm}
     \begin{center}
     \subfloat{\includegraphics[width=0.4\textwidth]{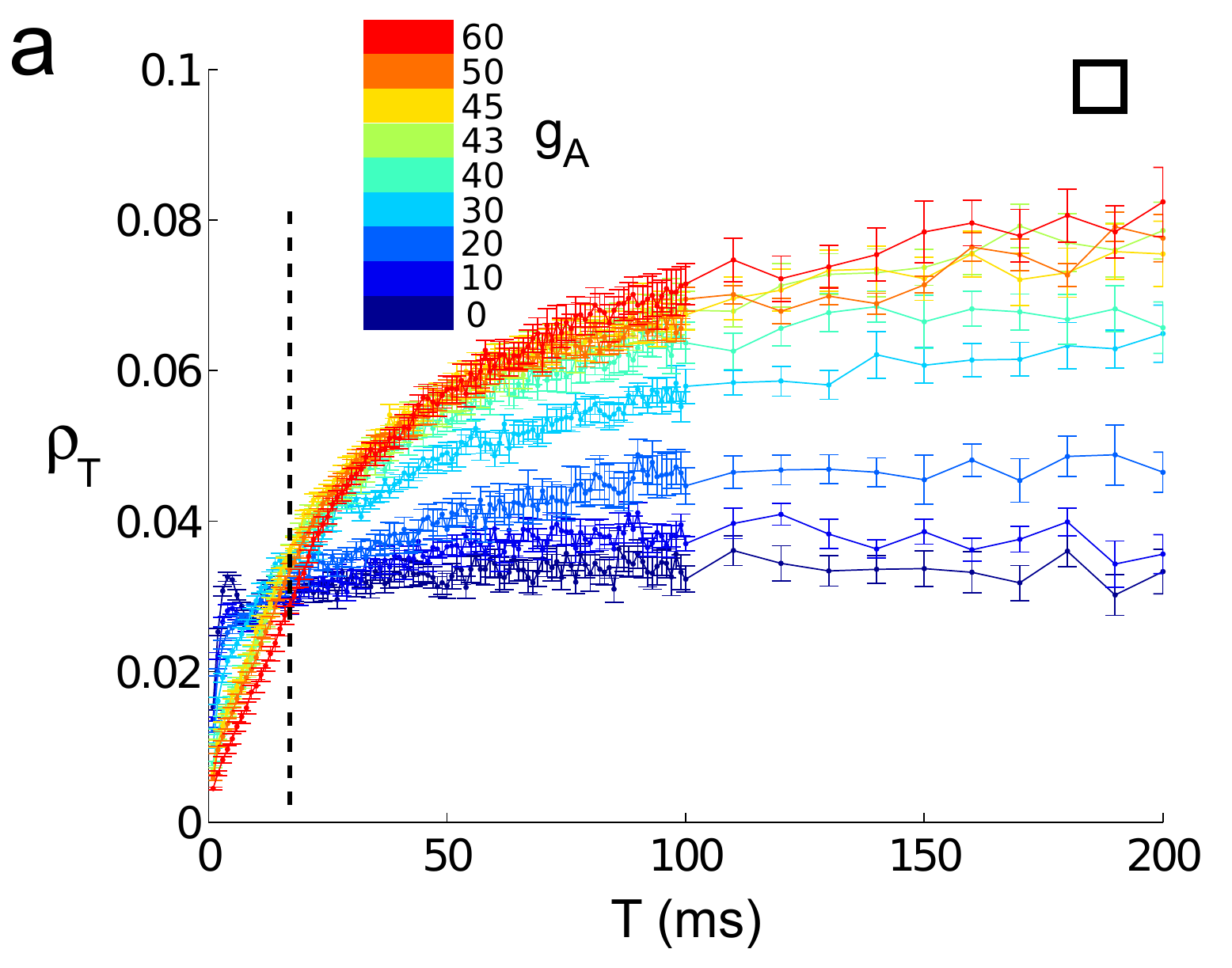}}\hspace{0.5cm}
     \subfloat{\includegraphics[width=0.4\textwidth]{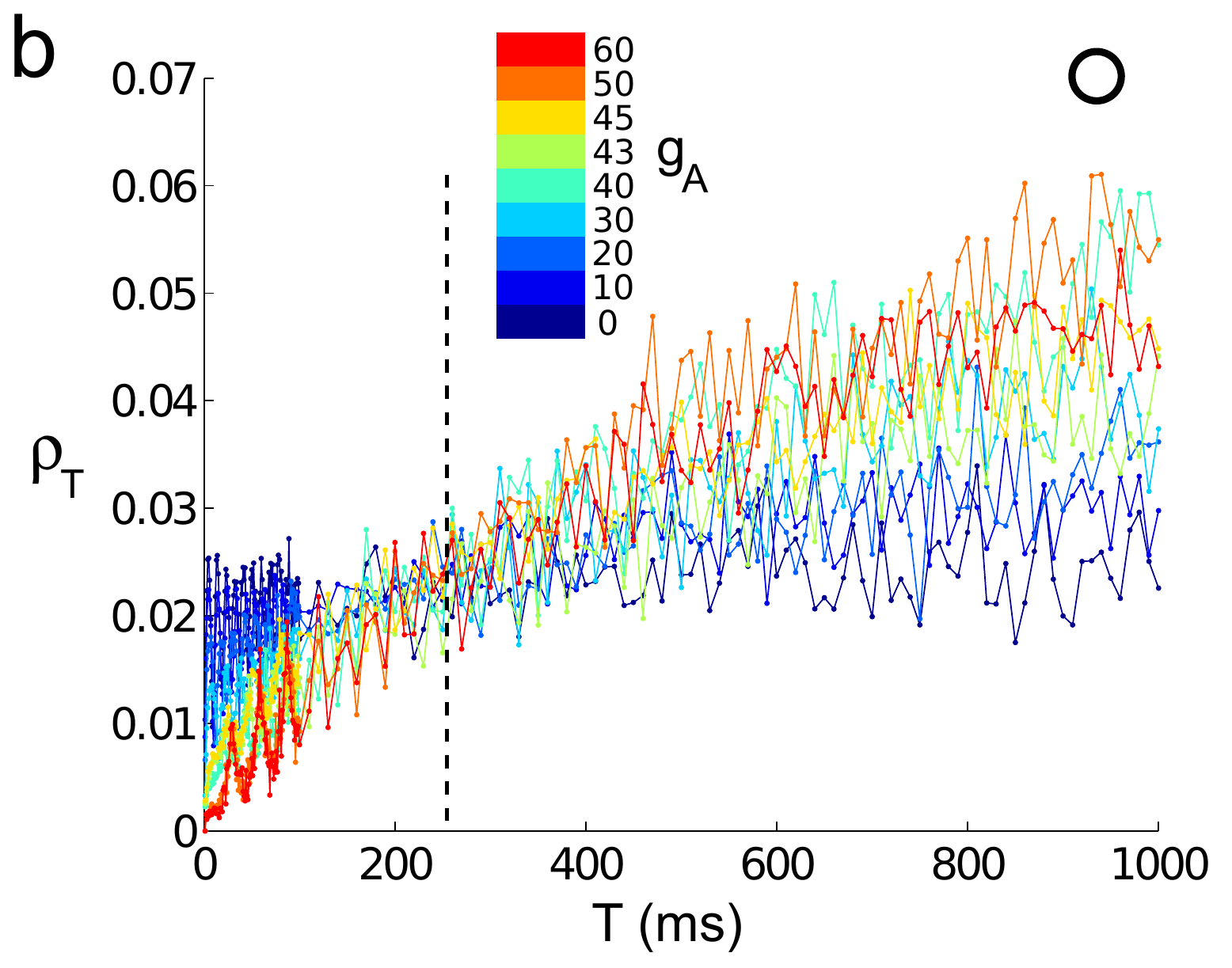}}\\
     \subfloat{\includegraphics[width=0.4\textwidth]{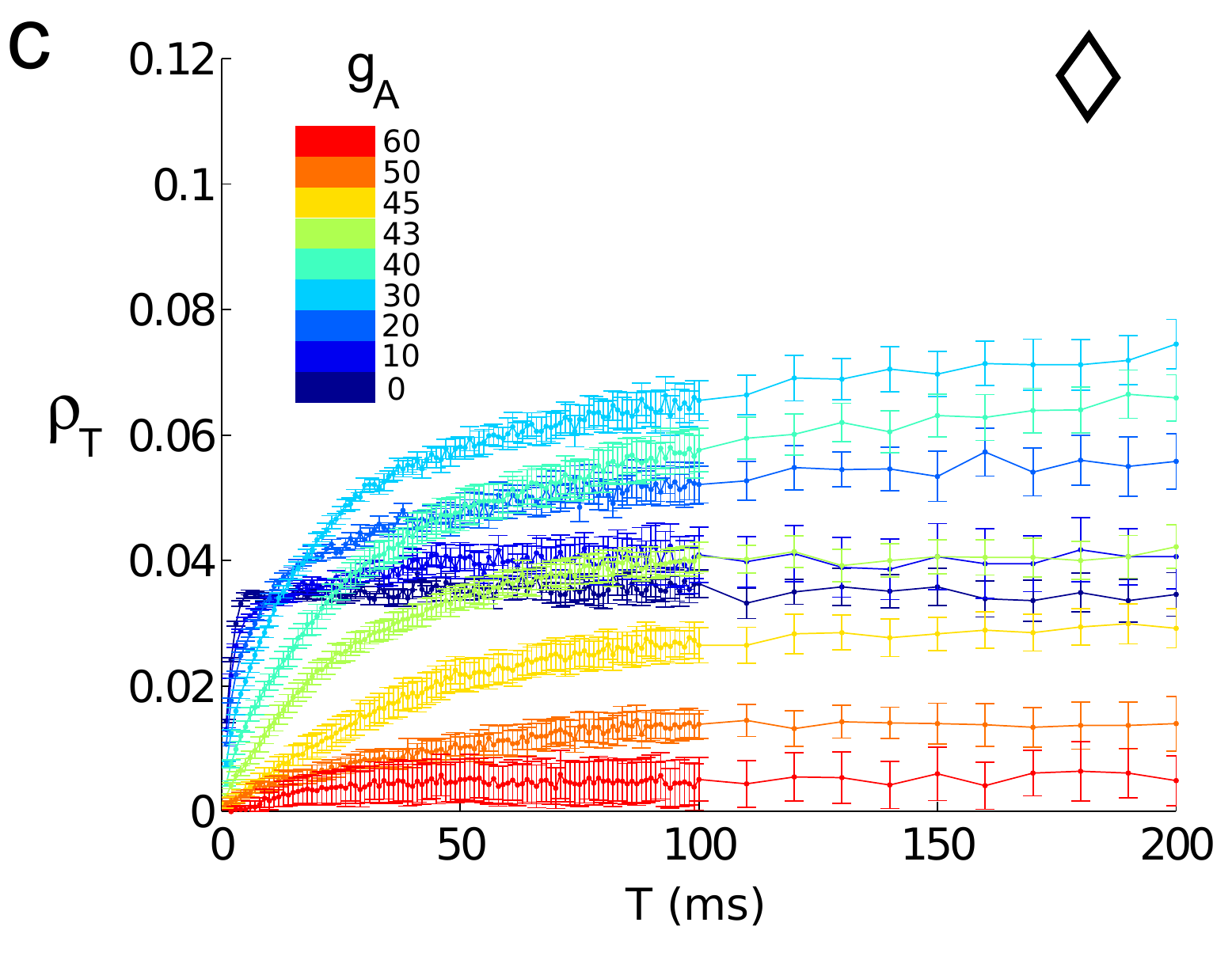}}\hspace{0.5cm}
     \subfloat{\includegraphics[width=0.4\textwidth]{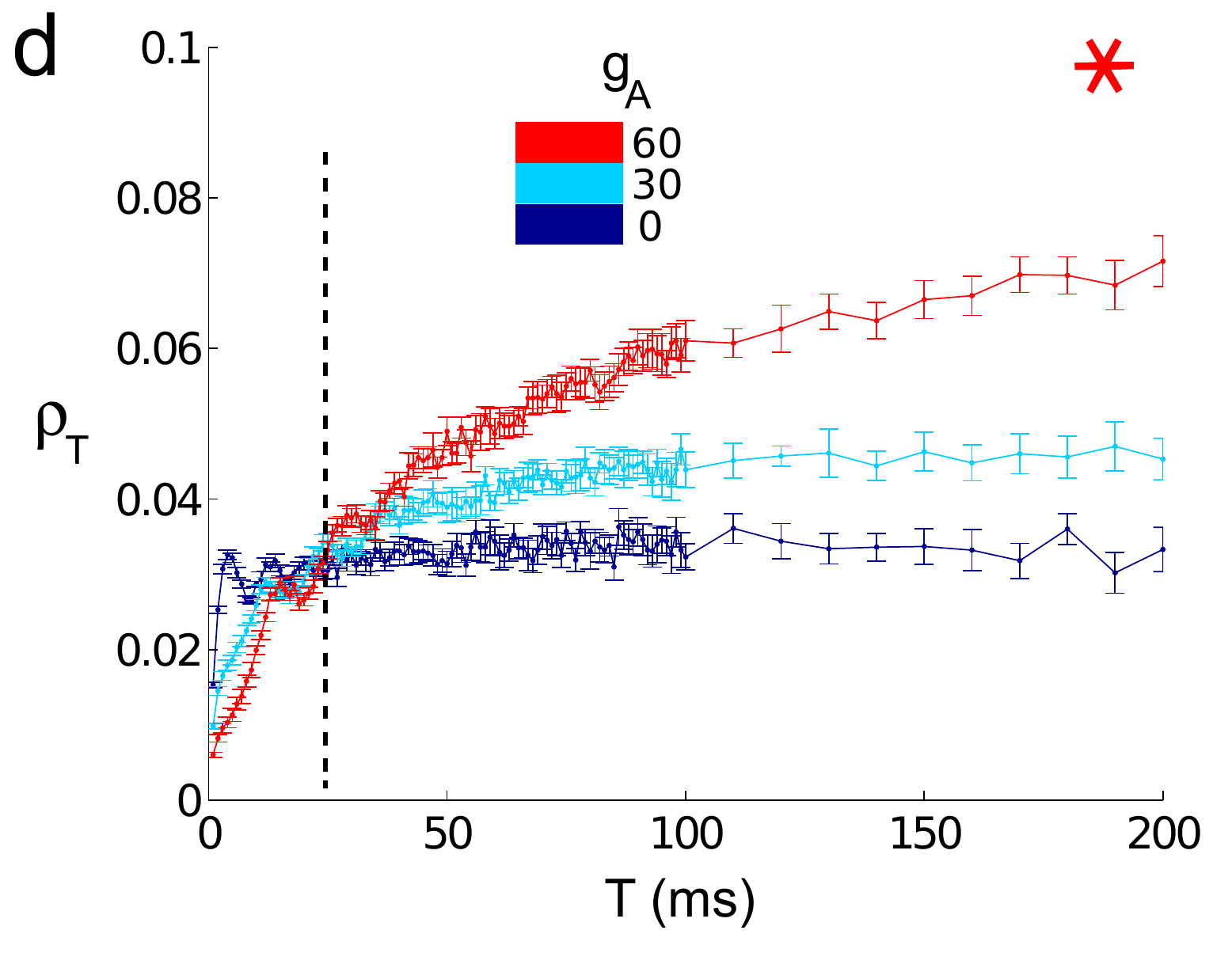}}
     \end{center}
\caption{Correlation coefficient $\rho_T$ vs. time window $T$. Colors indicate $g_A=0$ (dark blue) through $g_A=60$ (red) ${\rm mS}/{\rm cm}^2$.
The data from the superthreshold cases (a, b, and d) show the switch from Type II cells transferring more correlations to Type I cells transferring more as $T$ increases. The dotted line indicates the approximate time window $T_{switch}$ where the switch occurs. 
(a) High noise, superthreshold (b) Low noise, superthreshold, (c) Subthreshold, and (d) high noise, where input current $\mu$ has been chosen to match spike train variability across different values of $g_A$ (see text). }
 \label{fig:rho_vs_T}
 \end{figure}
 %%%%%%%%%%%%%%%%%%%%%%%%%%%%

We obtain additional insight into how spike count correlations depend on Type I vs. Type II spike generation and the time scale $T$ by choosing matched values of input parameters $\mu$ and $\sig$, and comparing spike count correlations produced for different values of the A-current conductance $g_A$.  We first concentrate on the $\mu$ and $\sigma$ values indicated by squares and circles in Fig.~\ref{fig:corrpanels}.  Both of these points indicate superthreshold inputs $\mu = I_{bif}(g_A)+2$ $({\rm \mu A}/{\rm cm^2})$ for all $g_A$ values.  The square corresponds to higher noise $\sigma=5$ ${\rm \mu A}/{\rm cm^2}$, and the circle to lower noise $\sigma=1$ ${\rm \mu A}/{\rm cm^2}$.  In Figure \ref{fig:rho_vs_T}, we plot $\rho_T$ for a full range of $T$ values from 1 to 200 ms, for nine values of $g_A$ between $g_A=0$ and $g_A=60$ (thus filling in intermediate values of $g_A$ and $T$ between those in Fig.~\ref{fig:corrpanels}).  For both superthreshold cases, we see that Type II neurons transfer more input correlation into output (spike) count correlation at small $T$, while Type I neurons transfer more at large $T$; this transition occurs, roughly, at a value $T_{switch}$ indicated by the dotted line.  We note that for the low noise case $\sigma = 1$ (Figure \ref{fig:rho_vs_T}(b)), the trends appear less ordered as $g_A$ varies; as we will see in the next section, this is because the cross-covariance function is more oscillatory here, so that $\rho_T$ has not yet converged to its asymptotic large $T$ value. 

%%%%%%%%%%%%%%%%%%%%%%%%%%%%%%%
\begin{table}
\begin{center}
	\begin{tabular} {| l | l |  c  |  c  |  c  |}
	\hline
	 Regime & Statistics  & $g_A = 0$ & $g_A = 30$ & $g_A = 60$ \\
	 \hline
	 \multirow{2}{*}{superthreshold, high $\sigma$ $\left( \square \right)$}  & $\nu$ (Hz) &  113.2  &   69.9 & 31.6  \\
	        & $FF_{T=256}$ &  0.059  & 0.0795   &  0.195 \\
	 \hline
	 \multirow{2}{*}{superthreshold, low $\sigma$ $\left( \bigcirc \right) $}  & $\nu$ (Hz) &  108.4  &  65.0  &  34.8 \\
	        & $FF_{T=256}$ &  0.0107 &  0.013  & 0.023   \\         	
	 \hline
	 \multirow{2}{*}{subthreshold $\left( \Diamond \right)$  } & $\nu$ (Hz) &  81.7  &  30.0  & 0.171  \\
	        &  $FF_{T=256}$ & 0.145  &  0.282  &  0.99  \\         	
	 \hline
	     superthreshold, & $\nu$ (Hz) &  113.2  &  82.5  &  68.5 \\
	        fixed variability ({\huge \textasteriskcentered})	& $FF_{T=256}$ & 0.059  &  0.059  &  0.059 \\       
         \hline  	
	\end{tabular}
\end{center}
\caption{Output firing statistics at each of the comparison points identified in Figure \ref{fig:corrpanels} (see text).
For each set of matched points, we note the firing rate $\nu$ and the Fano factor of spike counts over long time
windows (specifically, $T=256$ ms).} \label{table:outstats}
\end{table}
%%%%%%%%%%%%%%%%%%%%%%%%%%%%%%%

Subthreshold points, denoted by diamonds in  Fig.~\ref{fig:corrpanels}, were also compared:  the mean input current 
is chosen to be $\mu = I_{bif}(g_A)-1$ $({\rm \mu A}/{\rm cm^2})$, and the noise magnitude to be $\sigma=5$ ${\rm \mu A}/{\rm cm^2}$. Here, the differing dynamical
structure between Type II and Type I neurons is evident in the firing statistics (see Table \ref{table:outstats}): while the bistable Type II neuron ($g_A = 0$) 
sustains a substantial firing rate, the monostable Type I neuron ($g_A = 60$) barely fires at this level of input current. The
correlation coefficient $\rho_T$, is also very low for $g_A=60$ at all time windows (Figure \ref{fig:rho_vs_T}(c)); this is consistent with the relationship between correlation and firing rate identified in earlier studies~\cite{RochaDoironSJR07,SBJRD07}.  Overall, note that the correlation coefficient $\rho_T$ increases steadily with $T$ for the Type I neurons (high $g_A$) but stays roughly constant over a broad range of $T$ for Type II neurons (low $g_A$).  Thus, while we do not observe a clear value of $T_{switch}$ for all values of $g_A$ for the subthreshold point in Figure \ref{fig:rho_vs_T}(c), we see the same {\it relative} trends as for superthreshold points.  Below, we will see how this effect follows from filtering properties of Type I vs. Type II cells.

Because spike generation mechanisms vary widely as $g_A$ changes, the neuron models with matched input statistics at different values of $g_A$ in Figs.~\ref{fig:rho_vs_T}(a, b, c) do not all have the same firing  variability.  In the final panel of Fig. ~\ref{fig:rho_vs_T}, we address this by showing that the same trends in $\rho_T$ persist if we select values of $\mu$ to maintain constant firing variability for each value of $g_A$ (see Table \ref{table:outstats}; variability measured via large-time ($T= 256$ ms) Fano factor).  Here, we fix $\sigma = 5$ ${\rm \mu A}/{\rm cm}^2$; the required current value $\mu$ for each $g_A$ is indicated with a red star in Fig.~\ref{fig:corrpanels}.

In sum, for matched values of the mean and variance of input currents, a pair of superthreshold Type II (vs. Type I) neurons will produce greater spike count correlations $\rho_T$ at short time scales $T$.  For a wide range of choices for the mean and variance, there will be a value of $T_{switch}$ where this relationship reverses, so that Type I (vs. Type II) neurons produce greater $\rho_T$ for $T>T_{switch}$.  For matched subthreshold currents, similar trends are present; overall, the presence of a time $T_{switch}$ depends on how the input statistics are chosen.

\begin{figure}[t]
  \vspace{-.25cm}
     \begin{center}
     \subfloat{\label{fig:shortT}\includegraphics[width=0.4\textwidth]{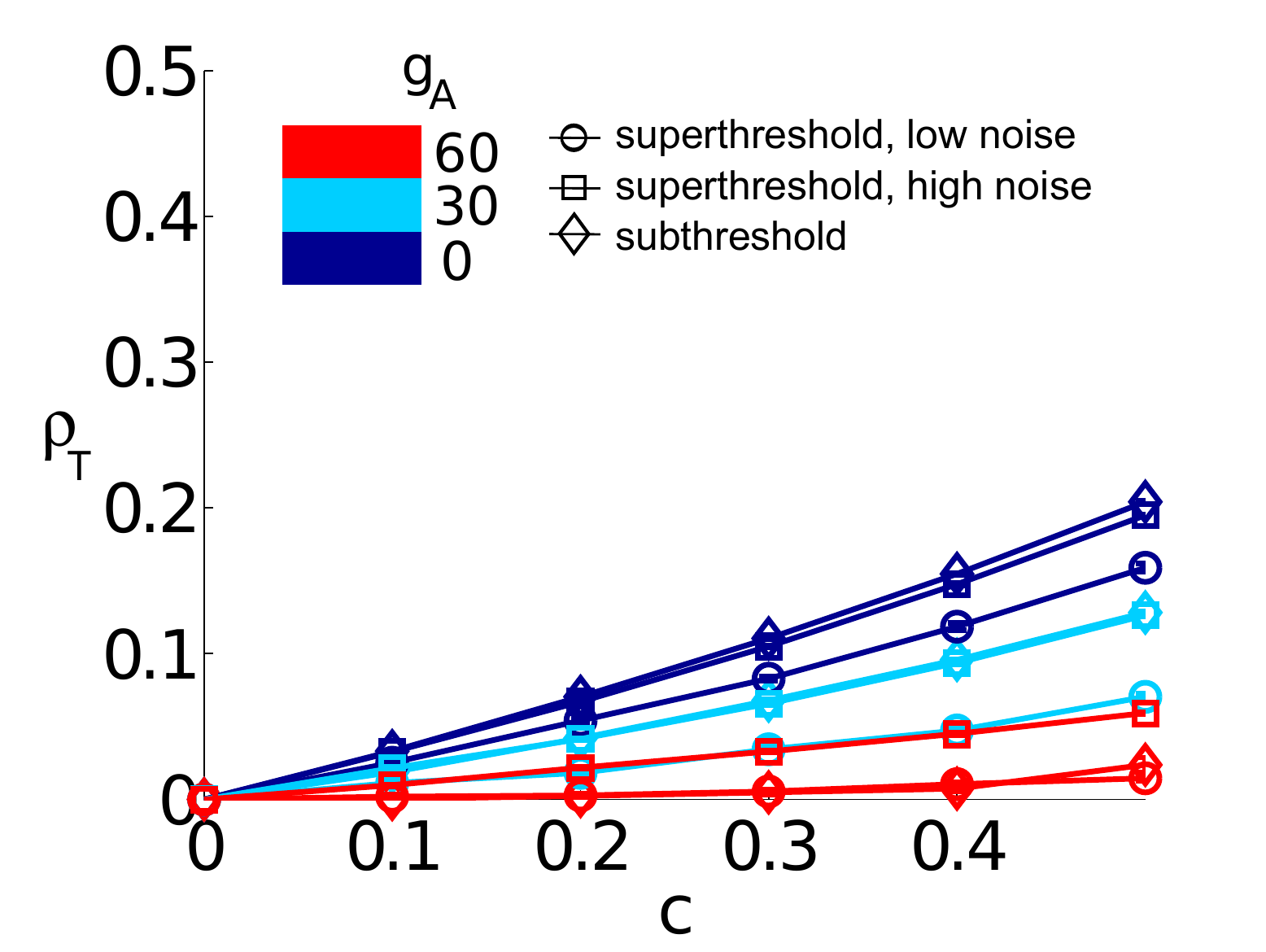}}
     \subfloat{\label{fig:longT}\includegraphics[width=0.4\textwidth]{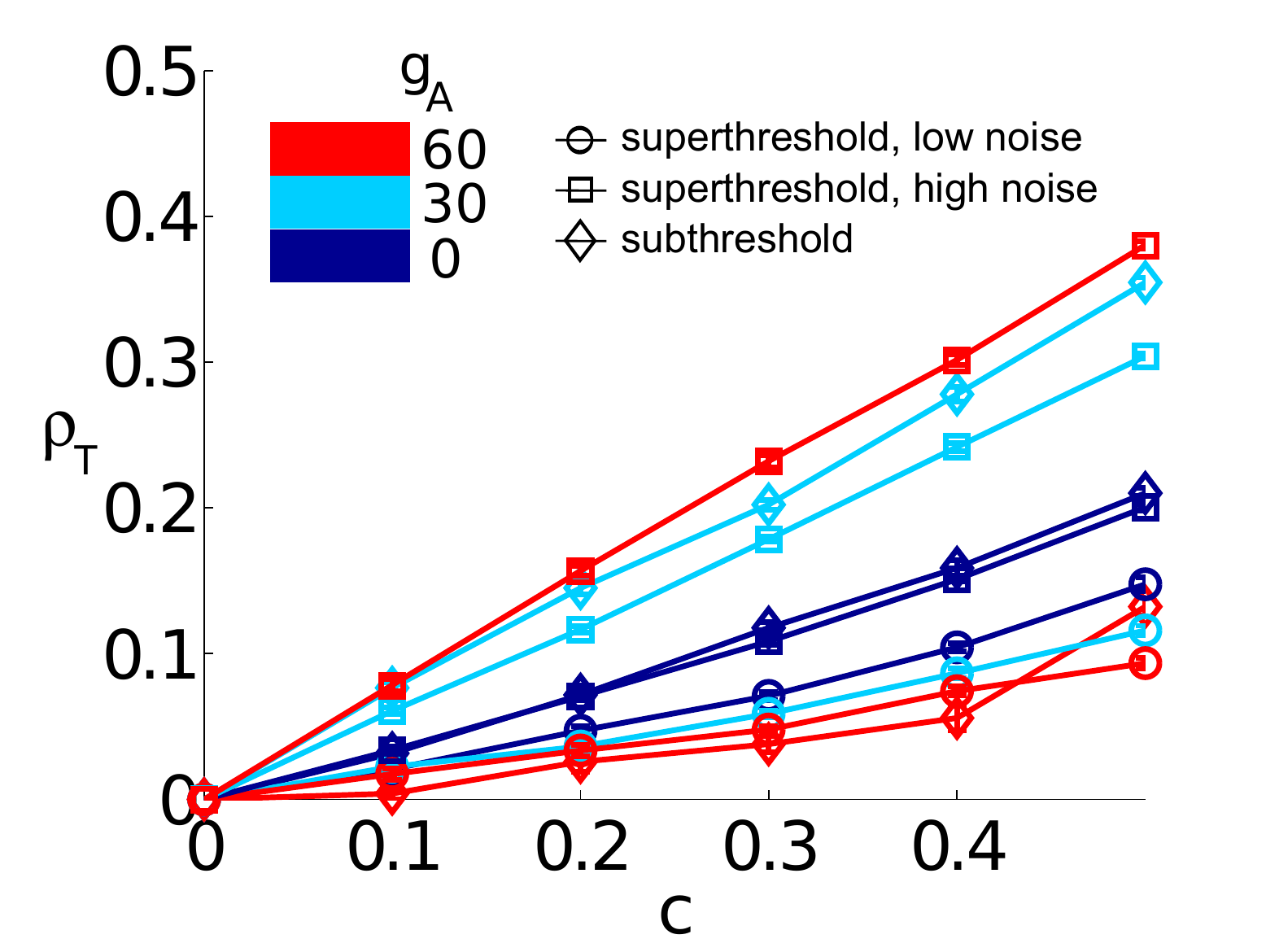}}
     \end{center}
\caption{Output correlation coefficient $\rho_T$ vs. input correlation coefficient $c$, showing an approximately linear relationship. Left: short time window ($T = 4$ ms). Right:  long time window ($T = 150$ ms).  Colors indicate $gA=0$ (dark blue),
$gA = 30$ (light blue), and $gA=60$ (red). Markers indicate relative location of $(\mu, \sigma)$-pair; subthreshold (diamond), superthreshold with low noise (circle), and superthreshold with high noise  (square).}
 \label{fig:rho_vs_c}
 \end{figure}

Finally, we note that the general trends observed here carry over, largely unchanged, to different values of $c$.  Figure \ref{fig:rho_vs_c} shows that, for the range $0.1<c<0.5$, trends in how $\rho_T$ changes with excitability type via ($g_A$) remain consistent.  In particular the relationship between input correlation $c$ and spike count correlation $\rho_T$ is roughly linear over this broad range of shared inputs.

%	\bei
%		\item {\bf TO DO: (J NSCI DOES NOT ALLOW SUPPLEMENTALS, SO WE'LL HAVE TO DO THIS IN MAIN TEXT: $\rho$ varies linearly with $c$. Show color plots of $\rho$ for three $gA$ values,
%		$gA=0,60$ and an intermediate point TBD. Show $c=0.1,0.3,0.5$. Probably large $T$)}
%		\eei

%%%%%%%%%%%%%%%%%%%%%%%%%%%%
%%%%%%%%%%%%%%%%%%%%%%%%%%%%

%%%%%%%
\subsection*{Trends in cross-correlation functions for Type I vs. Type II neurons}

The trends in spike count correlations that we have just described can be explained from the cross-covariance functions for neuron pairs, and how they differ as the characteristics of input currents and the level of the A-current conductance $g_A$ vary.  
We now demonstrate this via the cross-covariance functions shown in the right-most column of Fig.~\ref{fig:corrpanels}; these are for the superthreshold high noise cases ($\sigma = 5$) discussed in the previous section.  

To make the connection, recall that the spike count covariance, $\Cov(n_1,n_2)$, measured over a window of duration $T$ is given by the integral of
the cross-covariance function $C_{12}(\tau)$ against a triangular kernel of width $T$ (see Methods, and Equation~\ref{eqn:Cov_from_C}).  Thus, for short windows $T$, only the central peak of $C_{12}(\tau)$ contributes to spike count covariance.  In the limit of long windows $T \rightarrow \infty$, the spike count covariance is simply the integral of the cross-covariance function, multiplied by $T$, over the whole $\tau$ axis. Spike-count correlation $\rho_T$ is then given by the \textit{ratio} of spike count covariance to the spike count variance.  As we will show below, $\rho_T$ and $\frac{\Cov(n_1,n_2)}{T}$ often show similar trends with $T$.  Both quantities are of interest:  while $\rho_T$ gives a normalized metric of correlation, $\frac{\Cov(n_1,n_2)}{T}$ is the relevant quantity to analyze impact on downstream excitable cells (see below, ``Readout of correlated spiking by downstream cells").

\begin{figure}[p!]
  \vspace{-1cm}
     \begin{center}
     \subfloat{\includegraphics[width=0.4\textwidth]{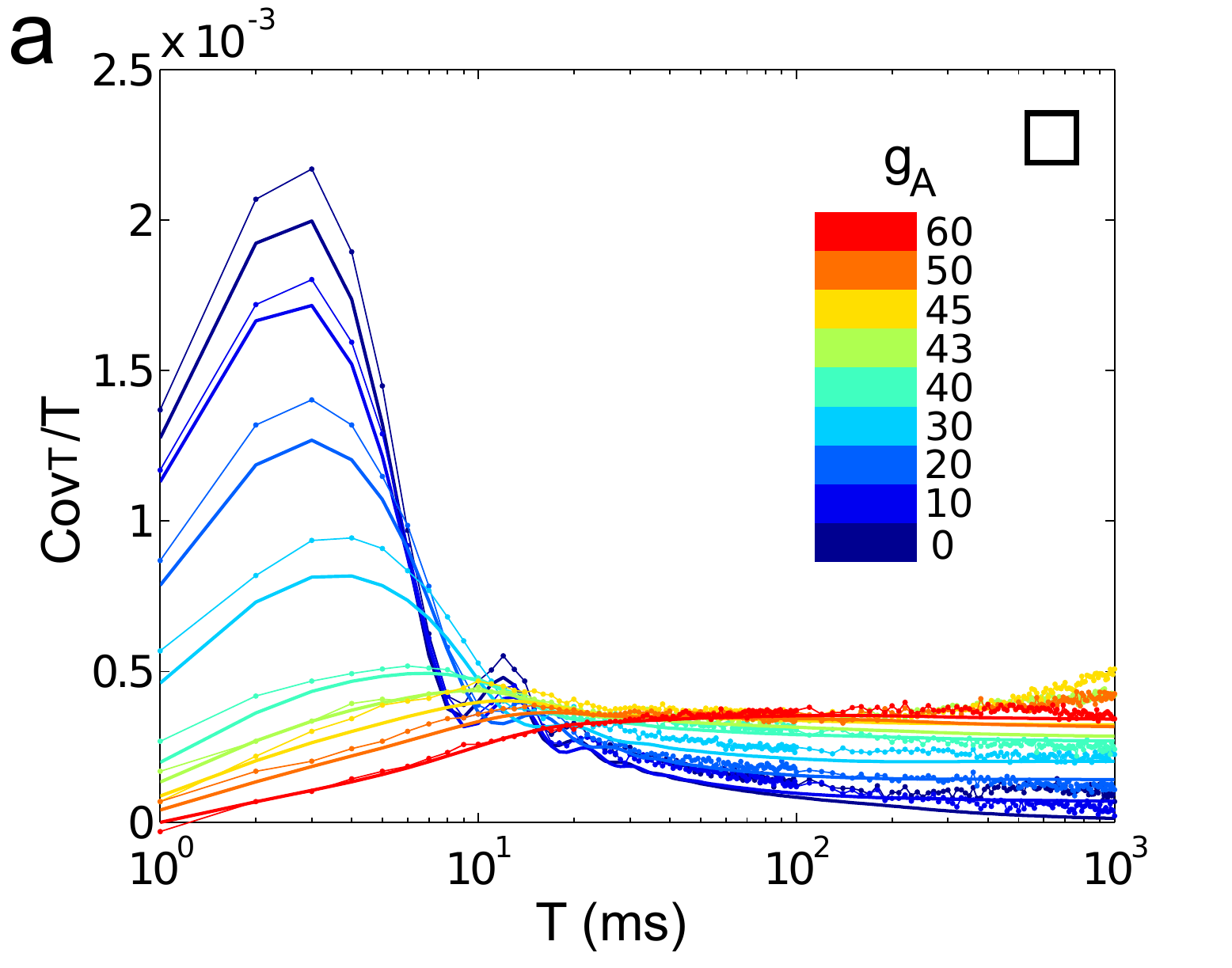}}
     \subfloat{\includegraphics[width=0.4\textwidth]{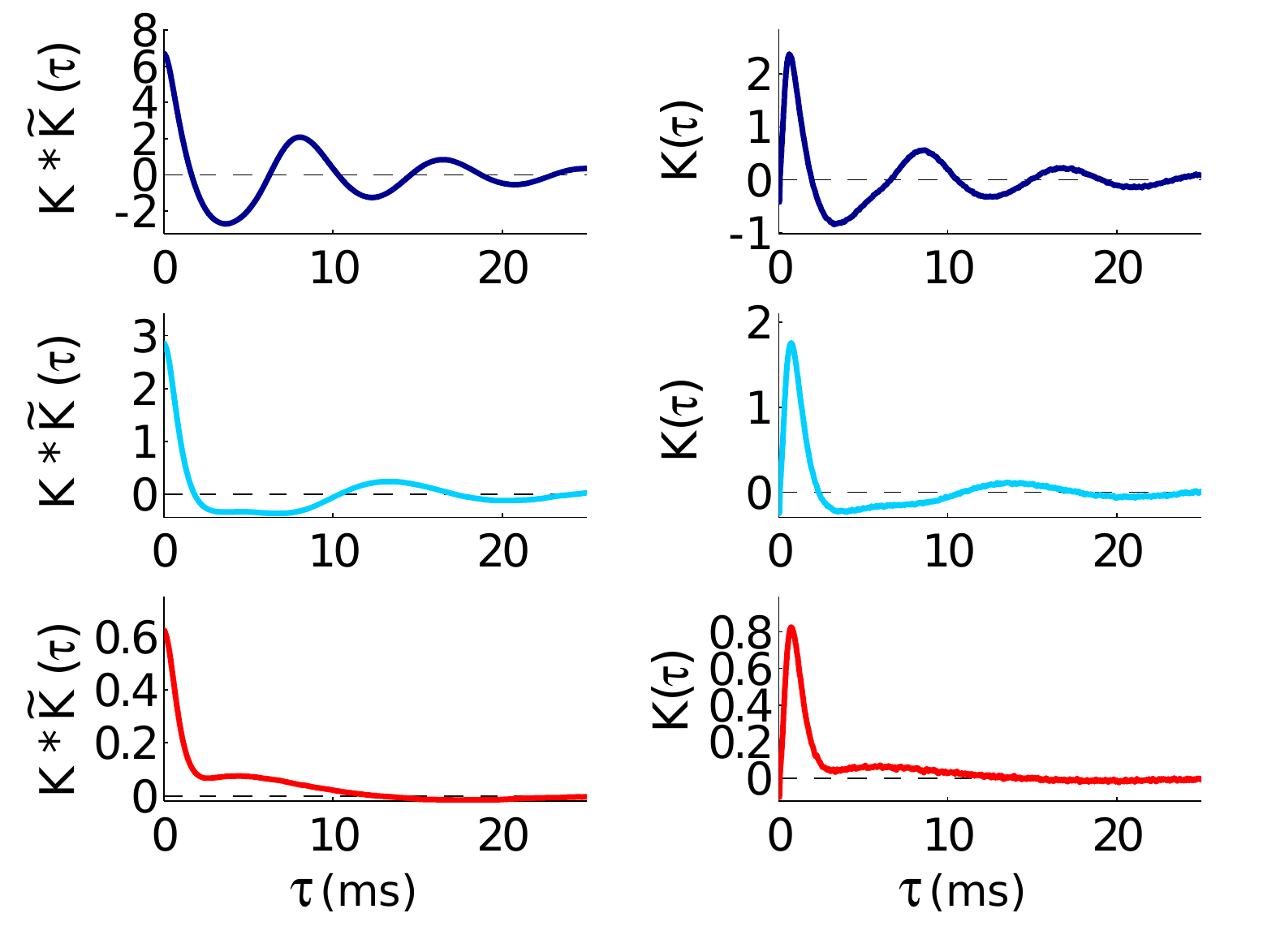}}\\
      \vspace{-.4cm}
       \subfloat{\includegraphics[width=0.4\textwidth]{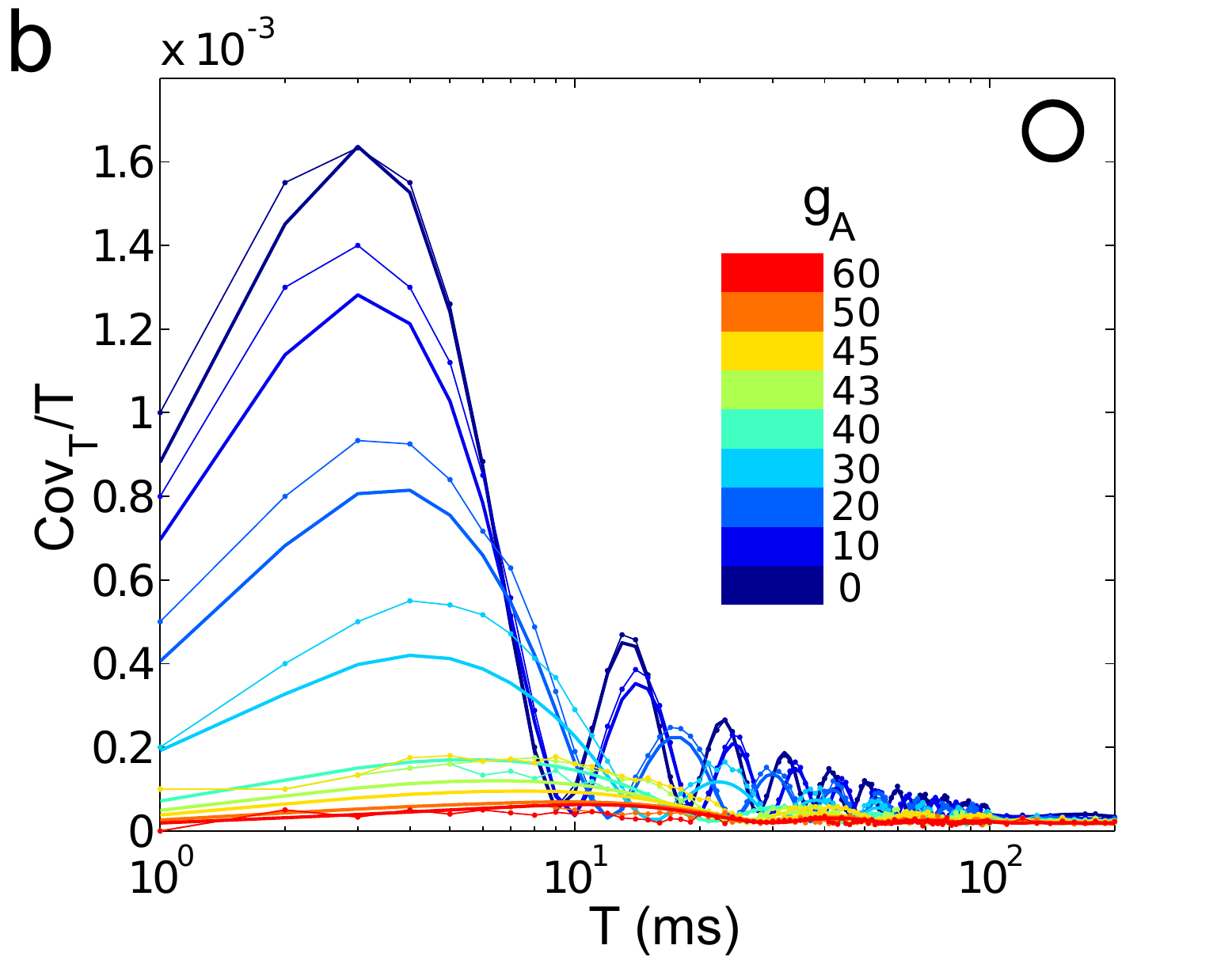}}
     \subfloat{\includegraphics[width=0.4\textwidth]{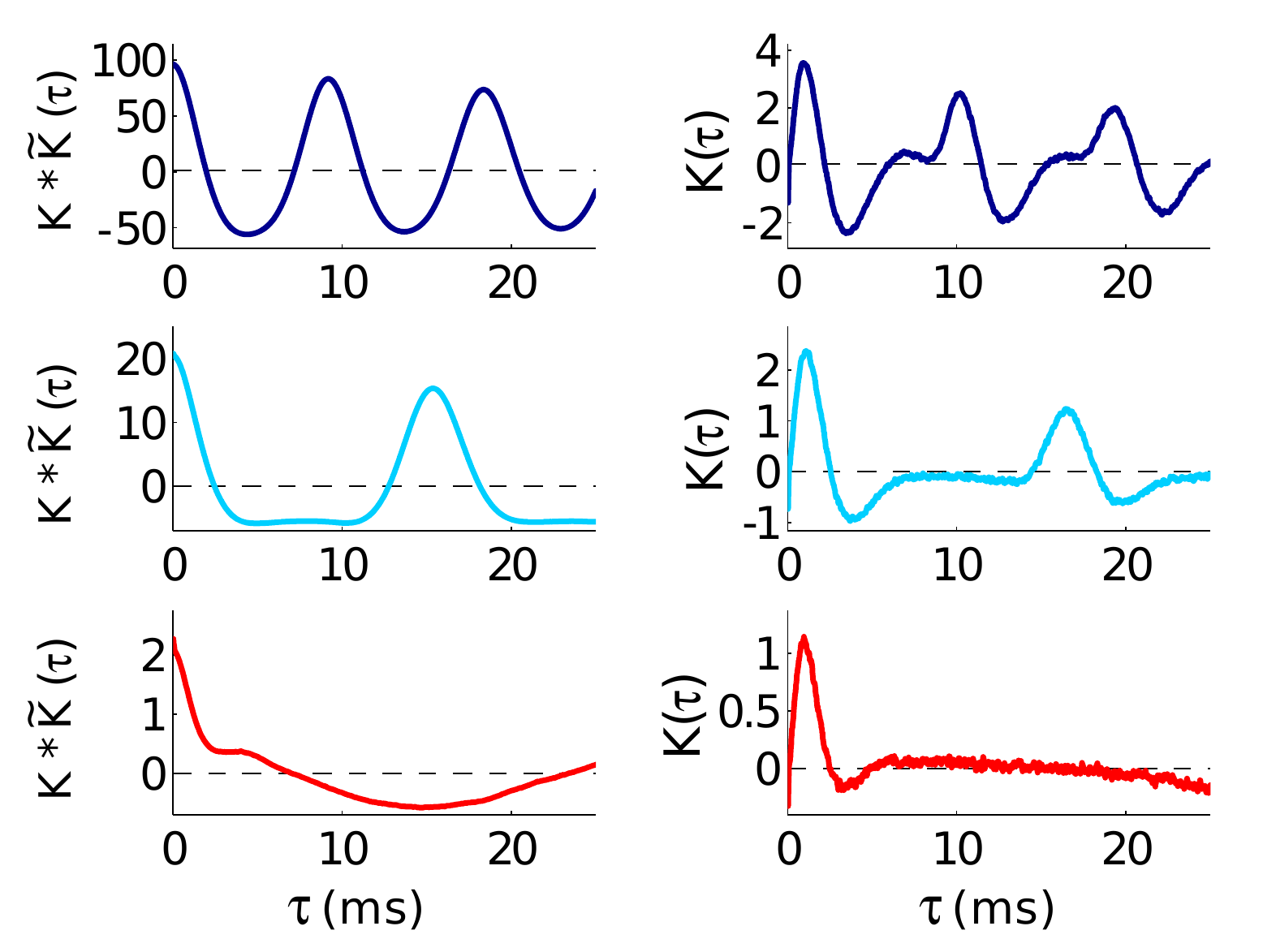}}\\
      \vspace{-.4cm}
     \subfloat{\includegraphics[width=0.4\textwidth]{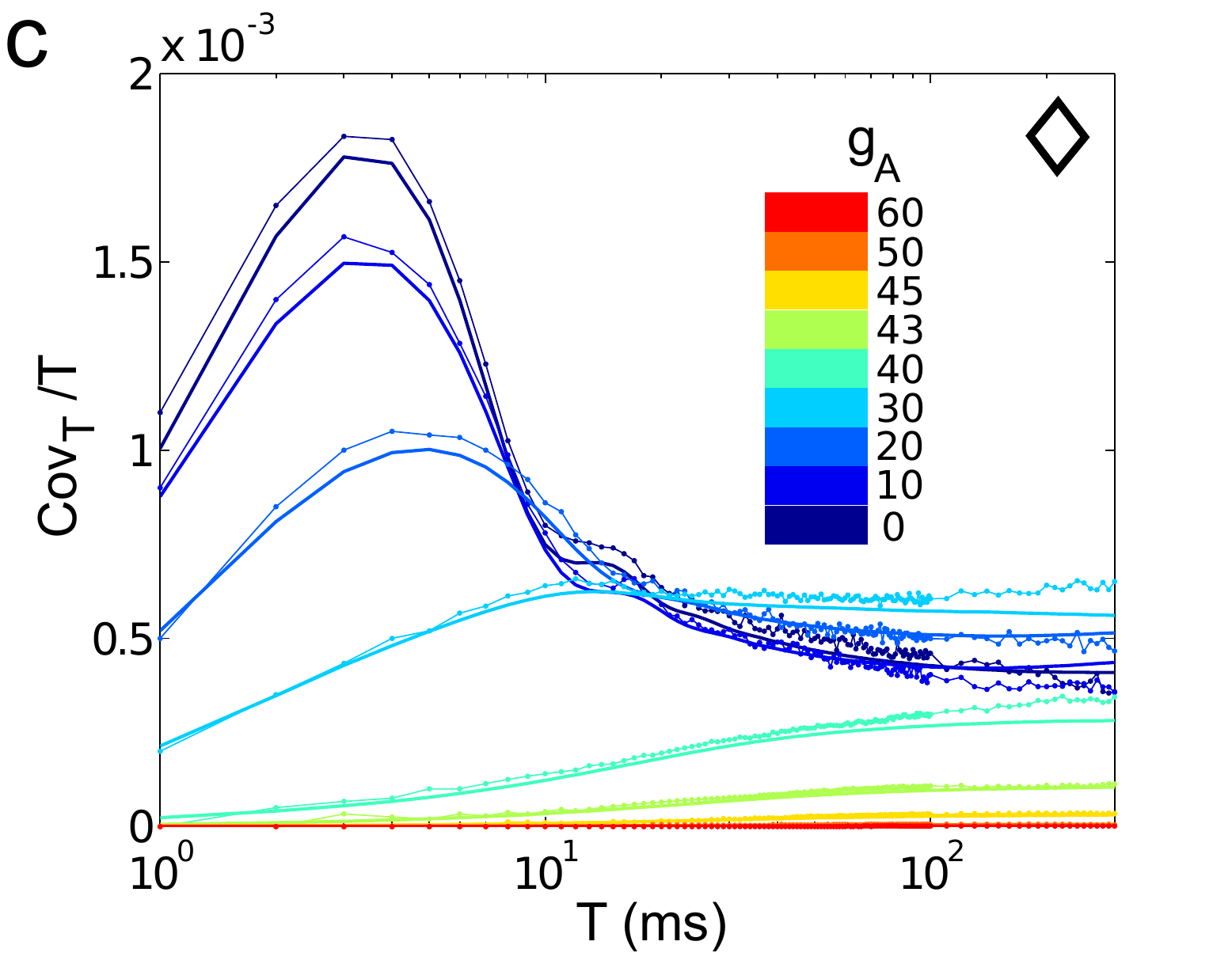}}
     \subfloat{\includegraphics[width=0.4\textwidth]{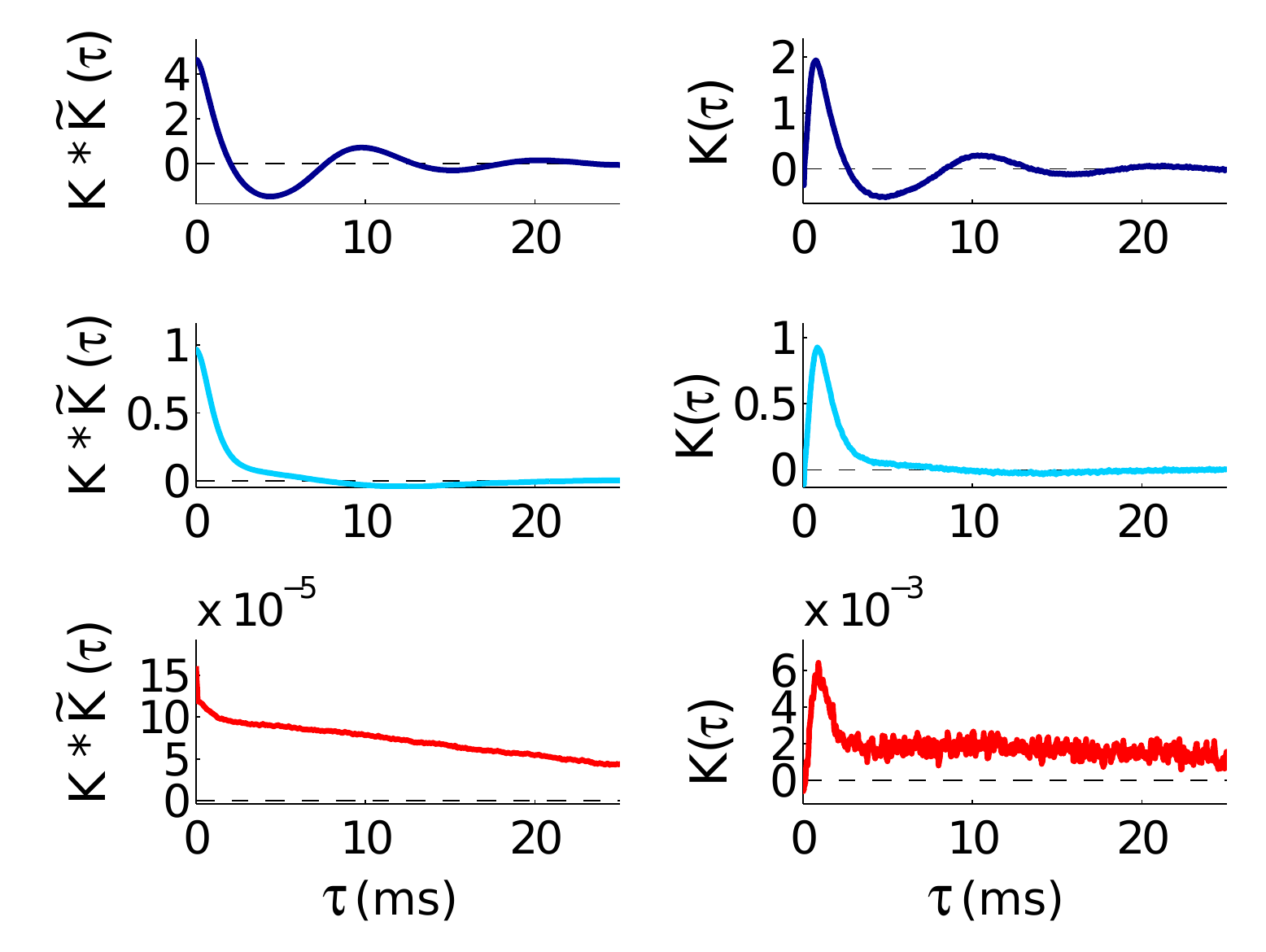}}\\
       \vspace{-.4cm}    
     \subfloat{\includegraphics[width=0.4\textwidth]{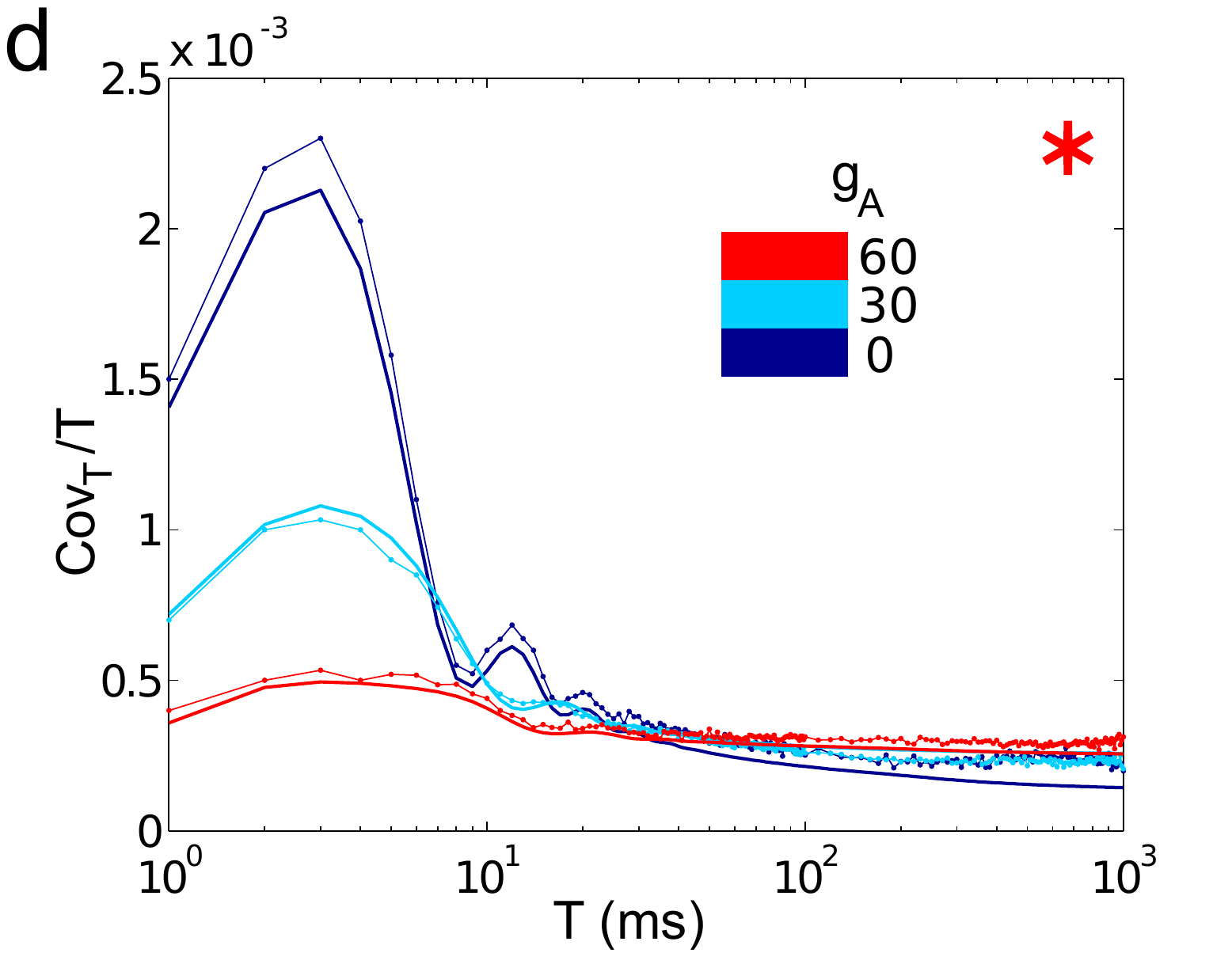}}
     \subfloat{\includegraphics[width=0.4\textwidth]{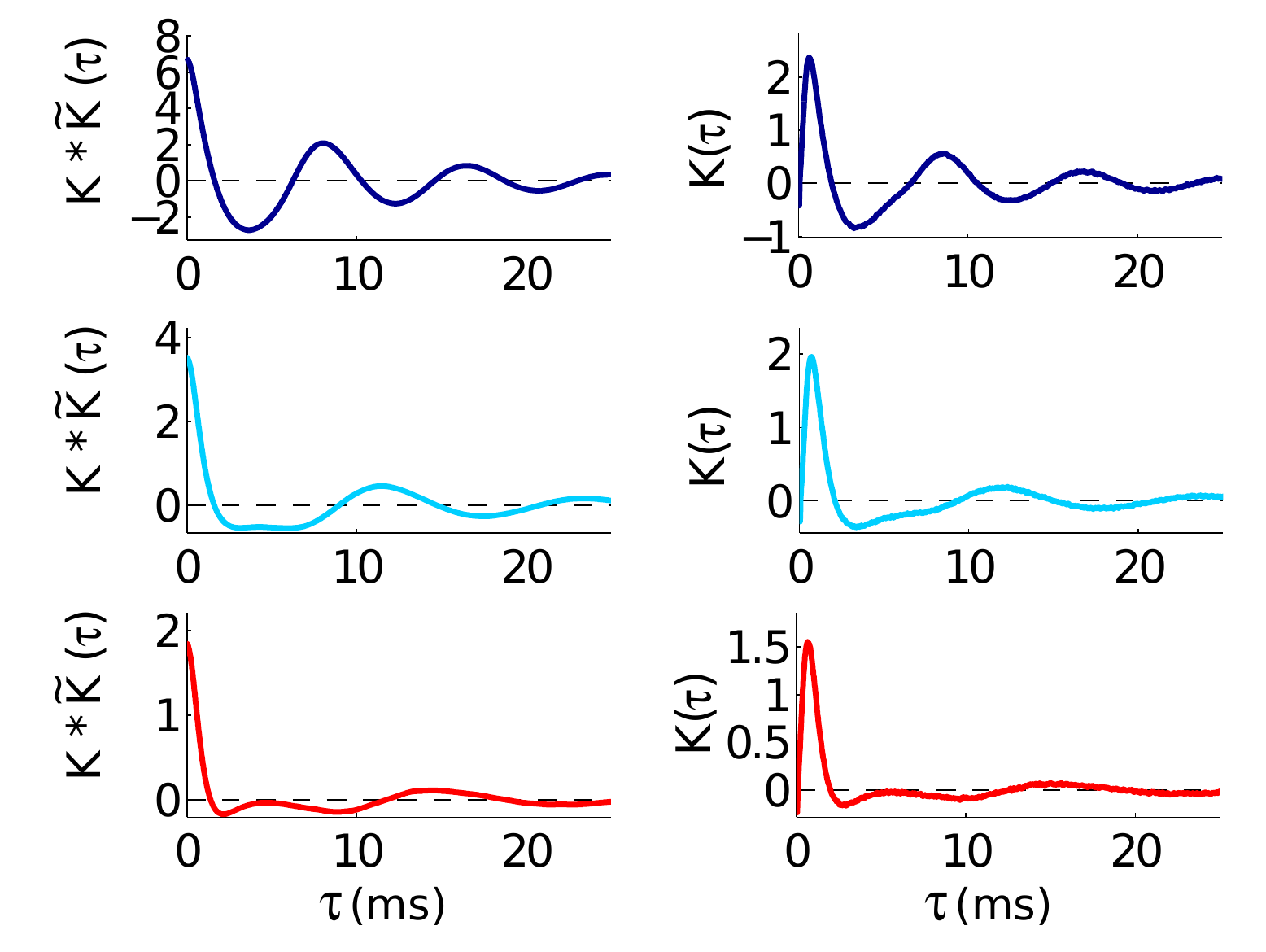}}
     \caption{Spike count covariances, and their relationship to spike triggered averages (STAs). Each row compares data collected at a comparison point for the input current statistics ($\mu$, $\sigma$); see text.  From top, superthreshold current with high noise (square),  superthreshold with low noise (circle), subthreshold (diamond), and superthreshold with high noise and matched variability (red star). (Left column) Actual (dotted lines) and predicted (heavy solid lines) spike count covariances ($\Cov(n_1,n_2)/T$), for representative points and all $g_A$ values. Colors identify $g_A$ values, which range from $gA=0$ (dark blue),
through $gA = 30$ (light blue),  to $gA=60$ (red) ${\rm mS}/{\rm cm}^2$. (Right column) Select spike-triggered averages (right column) and one-sided cross-covariance functions (left column, derived from the STA using Eqn.~\ref{eqn:C_from_K}) used to compute predicted spike count covariances. }\label{fig:Cov_Est_vs_Data}
\end{center}
 \end{figure}

Armed with these relationships between $C_{12}(\tau)$, ${\Cov(n_1,n_2)}$, and $\rho_T$, we revisit the trends observed in the previous section and explore their origin.  Starting in the upper-right of Fig.~\ref{fig:corrpanels}, note that $C_{12}(\tau)$ has a much larger central peak --- and hence short-T spike count covariance --- for Type II excitability ($g_A=0$) than for Type I (bottom-right, $g_A=60$).  This is also clear in Fig.~\ref{fig:Cov_Est_vs_Data}(a), where we plot spike count covariance vs. $T$.  Over long windows $T$, the trend reverses.  For $g_A = 0$, the cross-covariance function shows oscillations with significant negative and positive lobes.  These lobes tend to cancel as $C_{12}(\tau)$ is integrated over long windows $T$.  This cancellation results in little overall change in values of spike-count covariance computed at increasingly long values of $T$.
For Type I excitability, however, $C_{12}(\tau)$ is mostly positive, so that spike count covariance increases with $T$.  

As discussed above, spike-count correlation $\rho_T$ is given by the ratio of spike-count covariance and variance.  Comparing Fig.~\ref{fig:Cov_Est_vs_Data}(a) and Fig.~\ref{fig:rho_vs_T}(a) it is clear that spike-count correlation and spike-count covariance display the same trends as $g_A$ is varied.  For example, for very short times $T$ spike-count correlation $\rho_T$ is given by the ratio of the peak in $C_{12}(\tau)$ to that in $A_1(\tau)$; this ratio is also larger for Type II vs. Type I excitability.  

For other operating points ($\mu$, $\sigma$), while trends in spike-count correlation and spike-count covariance do not exactly agree, the some relative trends persist. % (compare Figs.~\ref{fig:Cov_Est_vs_Data}(b,d) and \ref{fig:rho_vs_T}(b,d)).  
Specifically, the general pattern that Type II cells produce greater covariance over short time windows, and that this trend disappears or reverses for larger time windows, holds for each of the superthreshold operating points explored here.  Overall, the major trends in spike count correlations stem from the presence vs. absence of large negative lobes in cross-covariance functions for Type II vs. Type I neurons.  We next describe how this difference arises via the distinct filtering properties of the two neuron types.

%%%%%%%%%%%%%%%%%%%%%%%%%%%%
% SPLIT FLOAT - CAPTION ON NEXT PAGE
 
%%%%%%%
\subsection*{Spike triggered average currents reliably predict spike count covariance}  \label{sec:explanation_STA}
%We next used the common input STA to estimate spike-count covariances at our reference points,
%using equations (\ref{eqn:C_from_K}, \ref{eqn:K_from_STA}).
The previous section showed how trends in spike count covariances for Type I vs. Type II neurons  follow from the presence of both strongly negative and positive lobes in cross-covariance functions for Type II neurons.  Here, we explain the origin of this phenomenon.  Equations~(\ref{eqn:C_from_K},\ref{eqn:K_from_STA}) (see Methods) provide the key link, in which the cross-covariance function is given in terms of a cell's spike-triggered average (STA), which is an estimate of the filter through which cells turn incoming currents into time-dependent spiking rates.  Here, we define the STA in response to common input as an average of the currents that precede spikes over a single long realization:
\begin{eqnarray}
\STA \equiv \left( \frac{1}{N}\sum_{k=1}^N I_c(t_k-\tau) \right),
\end{eqnarray}
where the $t_k$ are the $N$ spike times from the realization.
 
We first show that the prediction of spike count covariances from STAs is accurate. The left column Fig.~\ref{fig:Cov_Est_vs_Data} shows close agreement between spike count covariances computed from ``full" numerical simulation (heavy solid lines) vs. predictions from STAs via Eqn.~\eqref{eqn:C_from_K} (dotted lines).  
Next, we examine the shape of the STAs themselves (Fig.~\ref{fig:Cov_Est_vs_Data}, right column).  

For each operating point we consider, the Type II STA ($g_A = 0$) has a pronounced negative lobe.  Functionally, this corresponds to a ``differentiating" mode through which inputs are processed:  negative currents sufficiently far in the past tend to drive more vigorous spiking.  Biophysically, this corresponds to the kinetics of ionic currents, such as inward currents that can be de-inactivated through hyperpolarization.
By contrast, Type I STAs show a less prominent negative lobe, or none at all.  The resulting filtering of inputs is characterized as ``integrating:"  a purely positive filter is applied to past inputs to determine firing rate~(cf.~\cite{Day+01,ayafb03,ms08}).  

The consequences for spike cross-covariance functions are straightforward.  Trends are most pronounced for the superthreshold, high $\sigma$ case (Figure~\ref{fig:Cov_Est_vs_Data}(a)).  Here, the pronounced negative lobe in the Type II STA ($g_A = 0$) leads to a similar negative lobe in cross-covariance, and hence a sharp decrease --- following an initial
increase --- of spike count covariance as a function of time window $T$. The Type I STA ($g_A = 60$) is positive, leading to
a spike count covariance which steadily increases until it overtakes the Type II value at $T \approx 20$ ms.
These trends are also reflected in the spike count correlation $\rho_T$, as described previously.  Moreover, analogous plots for superthreshold, high $\sigma$ points, where spike-count variability (Fano factor) is maintained across $g_A$ values
(Fig.~\ref{fig:Cov_Est_vs_Data}(d)), show the same trends.

In the low $\sigma$ case (Fig.~\ref{fig:Cov_Est_vs_Data}(b)), the STAs have similar characteristics; 
however there is signifiant ringing in the STA at the characteristic frequency
of the oscillator. By the time the predicted (and actual) covariances $\frac{1}{T}\Cov(n_1,n_2)$ reach a limiting value, they are close to zero, and possibly too variable to
order definitively. It appears that covariance is still larger for Type II than for Type I at $T=200$ ms.
In the subthreshold case (Fig.~\ref{fig:Cov_Est_vs_Data}(c)), the STA for the Type I neurons is very small,
consistent with the very low firing rate here. The Type II neuron shows a more robust response, similar in magnitude but less oscillatory than
for the superthreshold regime.

\subsection*{Trends in spike-generating dynamics mirror trends in spike-triggered averages and transferred correlations}

%As described above, one prominent axis along which to classify neurons is in
%their response to injected current; two qualitatively different classes, I and II can be defined.
%Both types have the feature that as injected current increases, the system transitions from a stable rest state to periodic firing through a bifurcation.
%Type I neurons can sustain an arbitrarily low frequency as the applied current $I$ passes through $I_{bif}$; this can be produced by a saddle node on invariant circle (SNIC) bifurcation, in which two fixed points collide and disappear, producing a periodic orbit of infinite frequency.
%Type II neurons abruptly begin firing at a
%non-zero frequency as $I$ passes $I_{bif}$;  this may be produced by --- among other structures --- a subcritical Hopf bifurcation.
%
%\bei
%	\item Seems we can replace above para with one sentence + ref to Methods -- repeats I think allied Sec of Methods.
%\eei

The transition from Type II to Type I spike generation in the Connor-Stevens model --- as manifest in the progression from discontinuous to continuous spike-frequency vs. current curves in Fig.~\ref{fig:fIcurves} --- can also be characterized via the type of {\it bifurcation} that governs the transition from quiescence to periodic spiking as increasingly strong currents are injected (see Methods).

% (the original  parameter used was $g_A = 47.7$ \cite{connor77}).  
For $0<g_A < 46$ ${\rm m S}/{\rm cm}^2$, the transition occurs via a subcritical Hopf bifurcation, as voltage trajectories jump from a stable rest state to a pre-existing stable periodic orbit (limit cycle).  This transition is schematized in the upper-left cartoon in Fig.~\ref{fig:fIcurves}.    As this figure shows, for smaller values of $g_A$ in this range, the frequency of this cycle is high ($\gsim 60$ Hz).  The voltage-conductance dynamics near both stable structures -- the stable rest state and the limit cycle -- is oscillatory.  This creates a {\it resonator} property (see~\cite{izhikevich07} and references therein):  if they are properly timed, both negative and positive inputs cooperatively produce spikes or cause them to occur earlier than they would in the absence of inputs.  This is reflected in the negative and positive lobes in the STA $ \propto K(\tau)$ for the $g_A=0$ cases in Figure \ref{fig:Cov_Est_vs_Data}:  recall that the STA is the filter applied to incoming currents to determine firing rates.

By contrast, for large $g_A > 58$ ${\rm m S}/{\rm cm}^2$, a saddle-node on invariant circle (SNIC) bifurcation occurs (see Methods).  
As sketched in the upper-right cartoon in Fig.~\ref{fig:fIcurves}, in this case there is a pair of fixed points that form a ``barrier" to spike generation for subthreshold values of $\mu$, and the shadow, or ``ghost" of these fixed points still affects dynamics for superthreshold $\mu$ -- producing slow dynamics in their vicinity. 
The inputs that will elicit or accelerate spikes are those that will push trajectories past the fixed points, or their ghost, in a distinguished direction.  These inputs therefore tend to have a single (positive) sign.  This is referred to as an {\it integrator} property, and gives rise to the mostly positive STAs seen in Fig. \ref{fig:Cov_Est_vs_Data} for the $g_A=60$ cases.  (This argument breaks down for very low-variance (low $\sigma$) inputs, as we will see in the next section).

Between these two extremes in $g_A$, the minimum frequency in response to a ramp current decreases steadily, creating a gradual shift between Type I and Type II behavior. This gradual shift is mirrored in the neural dynamics, in which the slow regions in the state space become increasingly dominant.  This transition is clear in the spike triggered averages --- and therefore spike count covariances ---shown in Fig. \ref{fig:Cov_Est_vs_Data}. For example, for $g_A=30$, we find both distinctly ``Type II"-like and ``Type I"-like aspects in the high noise (Fig. \ref{fig:Cov_Est_vs_Data}(a)) and subthreshold (Fig. \ref{fig:Cov_Est_vs_Data}(c)) covariance trends respectively. In the former, spike count covariance increases --- then decreases --- with $T$; reflecting an oscillating $\STA$; the end result is that the (normalized) covariance 
at $T=200$ ms is lower than the covariance at $T=1$ ms. In the latter, the normalized covariance steadily increases with $T$, reflecting a non-negative $\STA$.

\subsection*{Phase response curves (PRCs) predict common-input STAs}
 \begin{figure}[p!]
  \vspace{-.25cm}
     \begin{center}
     \subfloat[Type I ($g_A = 60$)]{\includegraphics[width=0.4\textwidth]{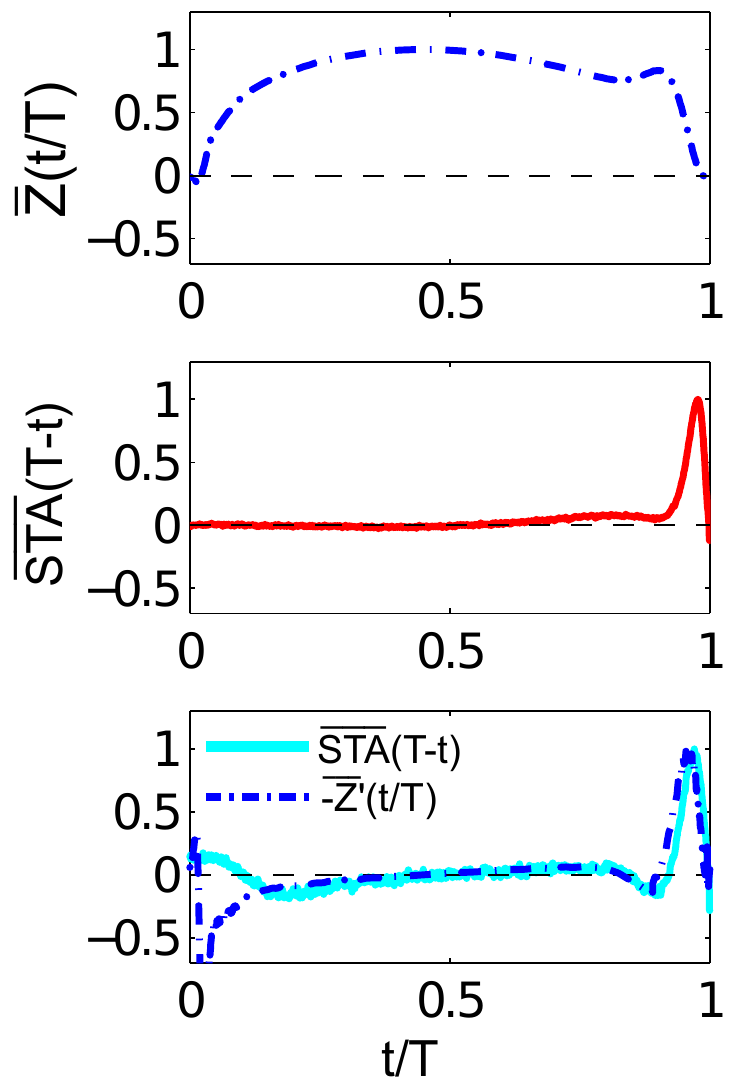}}
     \subfloat[Type II ($g_A = 0$)]{\includegraphics[width=0.4\textwidth]{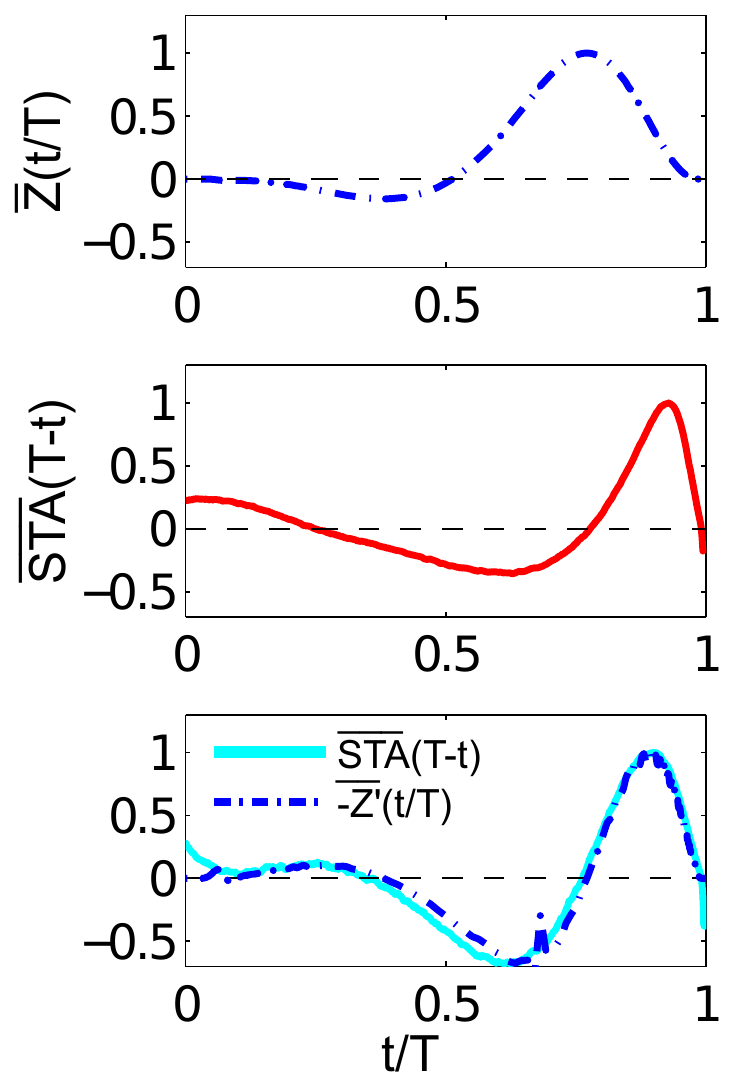}}\\
%     \subfloat{\includegraphics[width=0.4\textwidth]{dPRC_vs_STA_gA60.pdf}}
    % \subfloat{\includegraphics[width=0.4\textwidth]{dPRC_vs_STA_gA0.pdf}}
     \end{center}
\caption{Comparison of PRCs to spike triggered averages (STAs) computed for both low and high noise, for both Type I (left column)
and Type II (right column) neurons. For simplicity of visualization, each curve has been normalized by its maximum; 
that is $\overline{Z}(t)\equiv Z(t)/\max(Z(t))$,
$\overline{\STA}(t) \equiv \STA(t)/\max(\STA(t))$, and $\overline{-Z'}(t) \equiv -Z'(t)/\max(-Z'(t))$.
In addition, the time axis has been scaled by the mean period in each case.
Top row: PRC, showing monophasic and biphasic shape for Type I and Type II neurons respectively.
Middle row: High noise STA; the Type I neuron has lost the negative lobe in its STA, while the Type II neuron retains a negative component.
Bottom row:  Comparison of (blue dashed) dPRCs with STA for the low noise (cyan) case. Both STAs have negative components.}
 \label{fig:PRC_vs_STA}
 \end{figure}

  \begin{figure}[t]
  \vspace{-.25cm}
     \begin{center}
     \includegraphics[width=8cm]{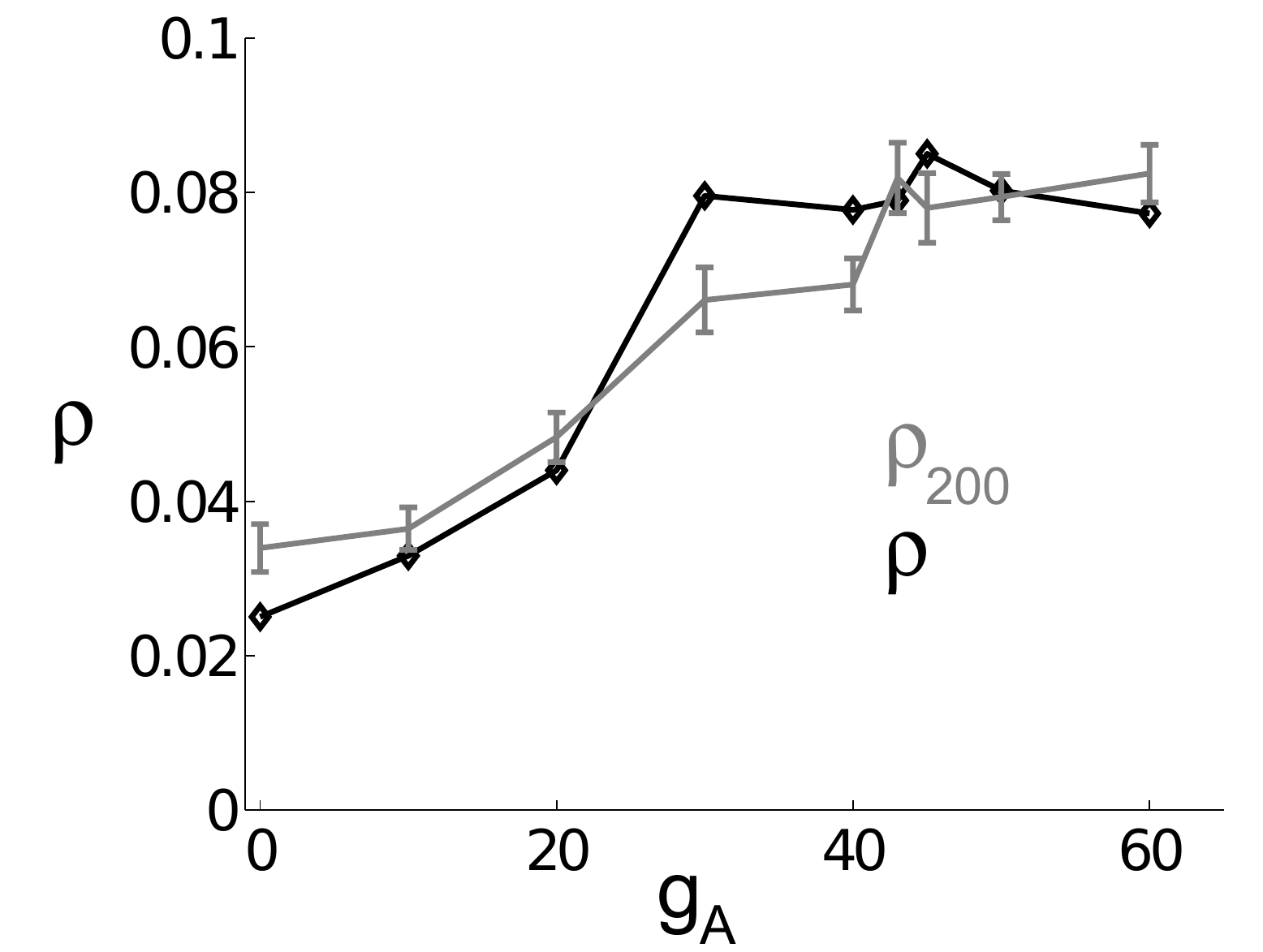}
     \end{center}
\caption{Correlation coefficient $\rho$ at time window $T=200$ ms, as $g_A$ is varied.
The data (gray solid) are from high-noise, superthreshold points and are the same as reported in
Figure \ref{fig:Cov_Est_vs_Data}.  The prediction (black solid with diamonds) uses Equation \ref{eqn:S_from_PRC}.
These data show the increase in long time scale correlation as the model transitions from Type II to Type I.}
 \label{fig:rho_data_vs_PRC}
\end{figure}

%------------------------------------------------

We next show how, for superthreshold operating points, the key properties of Type I vs. Type II spike generation that determine the filtering of common inputs can be understood via  
  a commonly used and analytically 
tractable reduced model for tonically spiking neurons.
%A neuron spiking tonically in response to a constant current injection can be thought of,
%dynamically, as traversing a limit cycle --- a periodic orbit --- in
%the governing dynamical system.
The response of such neurons to an additional small-amplitude current $I(t)$ can be
described by a \textit{phase model}, a one-dimensional description which keeps track only
of the progress of neuron along its periodic spiking orbit (or limit cycle).  Identifying progress along the cycle with a phase $\theta \in [0, 2\pi)$, this model is completely determined by a single function of phase $Z(\theta)$, called a \textit{phase response curve} or PRC \cite{ermentrout84,geomtime,ErmentroutTBook,ReyesF93}:
\begin{eqnarray}
\frac{d \theta}{dt} & = & \omega + Z(\theta) I(t).
\end{eqnarray}
We can interpret the meaning of this function by considering its effects on the timing of the next spike delivered
at a particular phase of the limit cycle $\phi$.
If $Z(\phi) > 0$, then a positive input delivered at that particular phase will push the neuron further along,
advancing the time of the next spike; if $Z(\phi) < 0$, the same input would delay the time of the next spike.

%Like the STA, the PRC can be related to dynamical features of a neuron.
Neurons that display Type I spiking have a purely positive (or Type I) PRC, while Type II neurons show a PRC that has both positive and negative lobes \cite{ermentrout84,ermentrout96,hansel95,BHMphase}. A purely positive PRC is characteristic of dynamics near a saddle node bifurcation, 
in which the system lingers near the ghost of its fixed points (as described in the previous section); input in a specific direction is needed to force the system away and elicit a spike. A biphasic PRC reflects oscillatory structure in the phase space, in which correctly timed negative and positive inputs can cooperate to elicit a spike (as with a Hopf bifurcation). 

Strong relationships between the PRC and the STA have been found for
neurons close to the threshold for periodic spiking (i.e., $\mu \gsim I_{bif}$, see Methods).
Spike-triggered covariance analysis of both a Type I phase model and the Wang-Busaki model show
that %, near spiking onset, 
the dominant  linear ``feature" (corresponding to the STA) qualitatively resembles the PRC \cite{ms08} in the presence of sufficient current noise. 
%Comparable Type II neurons show a more complex dependence on \textit{two} features,
%which may have significant power at frequencies corresponding to the firing rate and/or the subthreshold
%oscillation frequency \cite{ms08}. 
In the (Type II) Hodgkin-Huxley model, the two dominant ``spike-associated" features
identified through covariance analysis closely resemble the STA and its derivative; the STA, in turn, closely resembles
the PRC \cite{ayafb03}.

In contrast, phase models in the oscillatory regime (far from the excitability threshold)
are known to have an STA proportional to the \textit{derivative} of the PRC \cite{eguPRL07}.
We generalize this to the case of Fig.~\ref{fig:schematic}, where the relevant signal $\xi_c$ is delivered
\textit{on top of a noisy background} (see Eqn.~\eqref{eqn:STAPRC}, Methods): 
$\STA(t) \propto - Z'(-t)$.

In Figure \ref{fig:PRC_vs_STA}, we test the accuracy of these relationships for the superthreshold points considered above.
We show results for Type I ($g_A = 60$, left panels) and Type II neurons
($g_A = 0$, left panels), and compare the STA computed at two different noise levels to the
shape of the PRC ($Z(\theta)$) and its derivative, labeled dPRC ($Z'(\theta)$).
The time argument of the STA has been scaled so that one period ($T$) maps onto the unit interval;
likewise, the PRC is mapped onto the unit interval.
At the lower level of noise, we have good correspondence between the STA and the dPRC in both cases.  Notably, both Type I and Type II neurons have biphasic STAs.
At high noise levels, while there is not a strong \textit{quantitative} relationship between the STA and the PRC itself (unlike in the excitable regime
explored by \cite{ayafb03}), the PRC carries important clues about the \textit{qualitative} behavior of the STA.  The Type II neuron retains the
biphasic shape reflective of its PRC, while the Type I neuron has shifted to a purely positive STA.
In sum, by predicting the STA shape, the PRC gives important clues to the linear response
(and hence common input transfer) that we observe in Figure \ref{fig:Cov_Est_vs_Data}.

Finally, we test an alternate result (cf.~\cite{BarreiroST10,abouzeidArXiv})
that, in limited cases, relates PRCs to spike count correlations directly.  For
for long $T$ and reasonably small $\sigma$ and $c$,
\begin{eqnarray}
{\rho}  \approx & c \frac{\left[ \frac{1}{2\pi} \int_0^{2\pi} Z(x) \, dx \right]^2}{\frac{1}{2 \pi} \int_0^{2 \pi} \left( Z(x) \right)^2 \, dx} + O(\sigma^2) \,. \label{eqn:S_from_PRC}
\end{eqnarray}
In Figure \ref{fig:rho_data_vs_PRC}, we show that this gives a close approximation to simulation results for $T=200$ ms in the superthreshold, high-noise case (see Table 1).  Moreover, we can gain insight into the limitations this asymptotic approximation by comparing with the superthreshold, \textit{low}-noise case.  The results of~\citetext{BarreiroST10,abouzeidArXiv} are derived by taking the asymptotic limit
$T \rightarrow \infty$ \textit{before} considering $\sigma$ finite but small; in practice, the smaller the noise variance $\sigma$, the longer $T$ must be in order to see
this effect. For our low-noise points, the asymptotic behavior has not been recovered even at $T=1000$ ms 
(as may be seen in Figure \ref{fig:rho_vs_T}(b)). 
By using a very large (but probably biologically irrelevant) time window (data not shown), 
we eventually recover results consistent with the asymptotic prediction (Equation~\ref{eqn:S_from_PRC}).

%%%%%%%
\subsection*{Readout of correlated spiking by downstream cells} \label{sec:downstream}

%%%%%%%%%%%%%%%%%%%%%
 \begin{figure}[p!]
  \vspace{-.25cm}
     \begin{center}
       \subfloat{\includegraphics[width=0.4\textwidth]{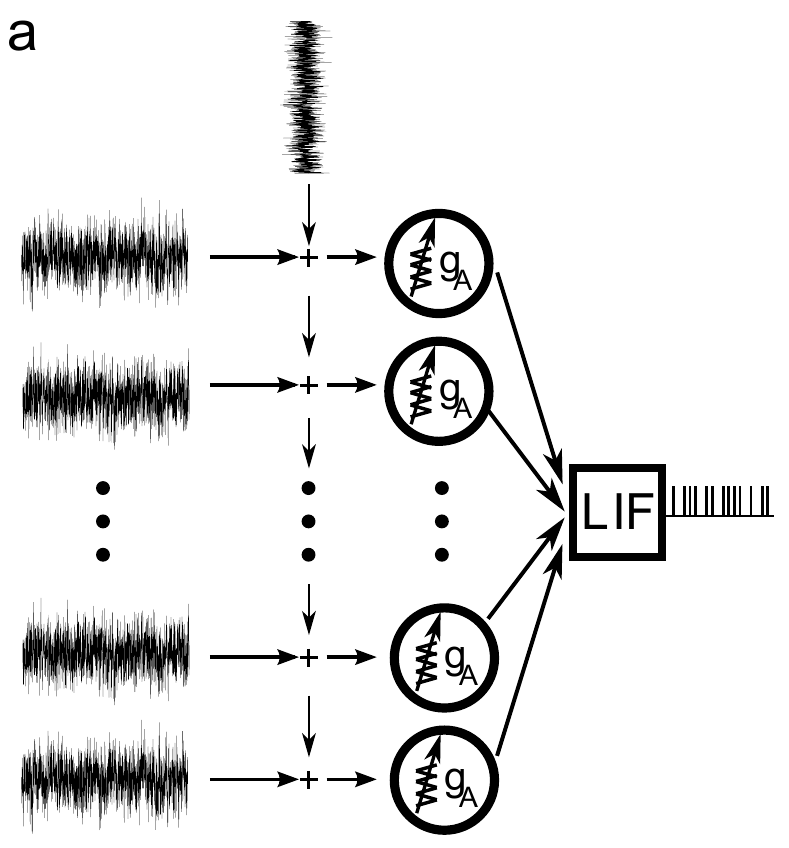}}
     \subfloat{\includegraphics[width=0.4\textwidth]{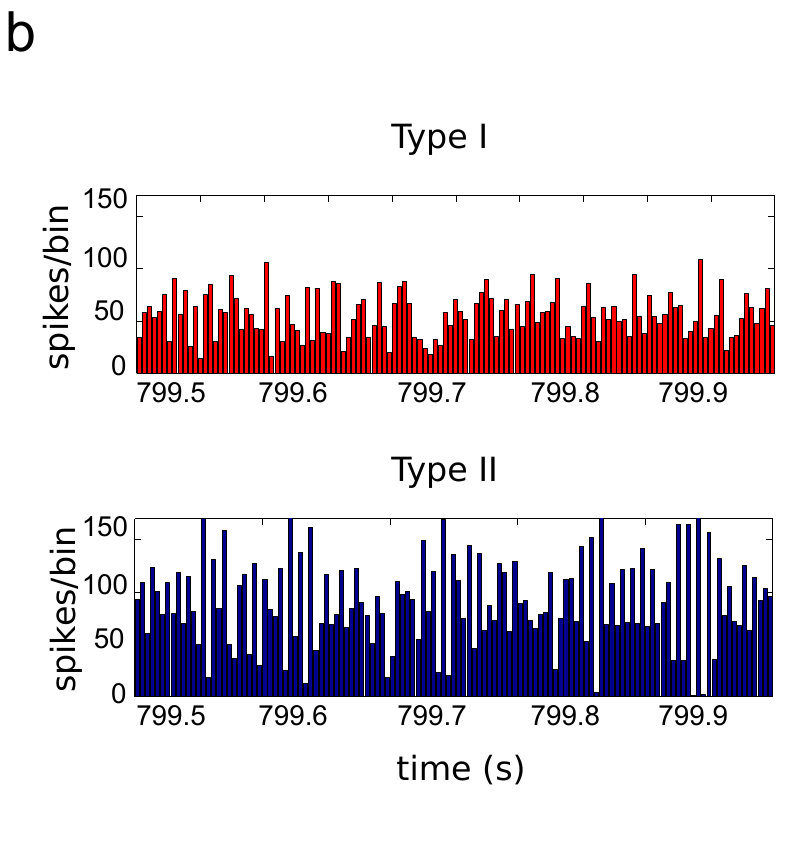}}\\
          \subfloat{\includegraphics[width=0.4\textwidth]{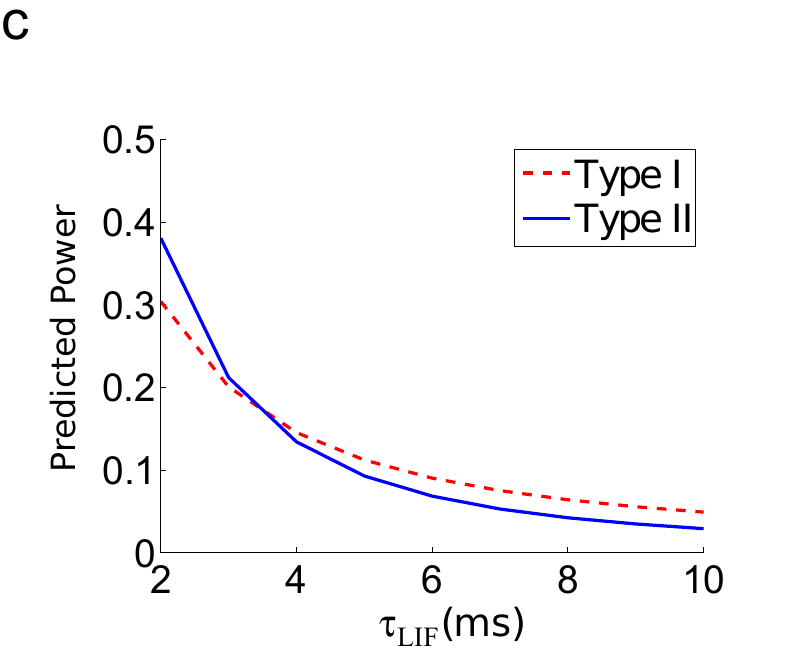}}
     \subfloat{\includegraphics[width=0.4\textwidth]{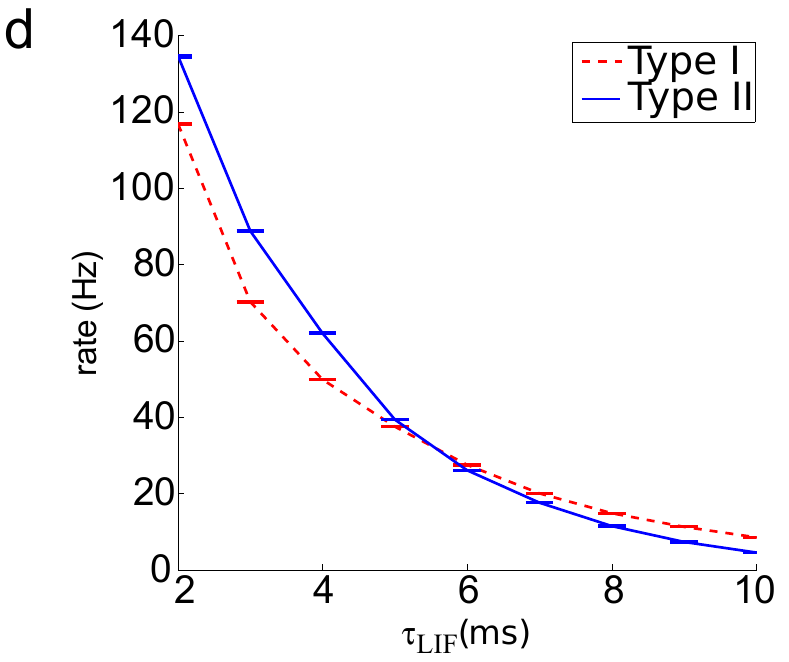}}
      \end{center}
 \caption{  (a) Schematic of ``upstream" Type I or Type II neuron population receiving common and independent inputs, and converging to a Leaky Integrate and Fire (LIF) cell downstream.
 (b) PSTHs from Type I and Type II upstream populations. (c) Predicted power of the voltage fluctuations in the LIF cell, using STA (see text). (d) Actual firing rates of the LIF cell, showing similar trends with LIF time scale $\tau_{LIF}$.} \label{fig:LIF_spike_rate}
\end{figure}
%%%%%%%%%%%%%%%%%%%%%

How could the difference in spike count correlation between Type I and Type II cells impact neural circuits?  We explore this impact in a simple network, in which correlated Type I or Type II cells collectively converge to drive a neuron downstream (see Fig.~\ref{fig:LIF_spike_rate}(a)).  

In more detail, the drive comes from a population of $N=200$ identical Type I ($g_A = 60$ ${\rm mS}/{\rm cm}^2$) or Type II ($g_A = 0$ ${\rm mS}/{\rm cm}^2$) upstream neurons; we refer to these as population I and population II respectively.  The upstream populations receive correlated inputs with $c=0.5$ and values of $\mu$ and $\sigma$ that yield matched levels of variability, as for the parameter set identified with stars in Fig.~\ref{fig:corrpanels} (for population I, $\mu = 18$ ${\rm \mu A}/{\rm cm^2}$ and $\sigma = 5$ ${\rm \mu A}/{\rm cm^2}$; population II, $\mu =-6 $ ${\rm \mu A}/{\rm cm^2}$ and $\sigma = 5$ ${\rm \mu A}/{\rm cm^2}$). This yields firing rates of 
$\nu_I = 63.5$ Hz for neurons in population I, and $\nu_{II} = 113$ Hz in population II.
Each upstream neuron has a single, instantaneous (delta function) synapse onto the downstream neuron of strength $g_{I}$ or $g_{II}$;
the relative size of the EPSPs are chosen so that the mean driving current is equal for each population ($\nu_I g_I = \nu_{II} g_{II}$, so that $g_I = 0.825$ mV, $g_{II} = 0.5$ mV).
%, and so that the LIF neuron is fluctuation-driven.

The total input received by the downstream neuron, $I_{ds}$, is thus the weighted sum of $N$ upstream spike trains $y_j(t)$:
\beq
I_{ds}(t) = g_I\sum_j y_j (t) \; \;  \; \mbox{   or      } \; \; \; I_{ds}(t) = g_{II} \sum_j y_j (t) \,.
\eeq
When the population size $N$ is large, the summed signal has the same temporal characteristics as the the cross-covariance between neuron pairs.
% signal has power at the same frequencies at which correlation is enhanced between driving neuron
%pairs:
Specifically, the autocovariance of the summed input is
\begin{eqnarray}
A_{ds}(\tau) & = & \Ex \left[ \left( I_{ds}(t) - \langle I_{ds} \rangle \right) \left(  I_{ds}(t+\tau) - \langle I_{ds} \rangle  \right) \right] \\
%& =&  \Ex \left[ \sum_i (y_i (t) - \nu_i) \sum_j (y_j (t+\tau) - \nu_j) \right] \nonumber \\
& = & N(N-1) \Ex \left[ (y_i(t)- \nu_i) (y_j(t + \tau)- \nu_j) \right] + N \Ex \left[ (y_i(t)- \nu_i)(y_i(t+\tau)- \nu_i) \right] \nonumber \\
& \approx & N^2 \Ex \left[ (y_i(t)- \nu_i) (y_j(t + \tau)- \nu_j) \right] \nonumber \\
& = & N^2 C_{12} (\tau)  \;.  \label{eqn:A_from_C12}
\end{eqnarray}
This relationship is evident in the peri-stimulus time histograms (PSTHs) in Fig.~\ref{fig:LIF_spike_rate}(b).  For population II, fast fluctuations above and below the mean population output reflect the negative lobe in $C_{12}(\tau)$ adjacent to its large peak.  Meanwhile, fluctuations in the output of population I are less extreme and more gradual in time.  

The downstream cell integrates $I_{ds}(t)$ via Leaky Integrate-and-Fire (LIF) voltage dynamics~\cite{Day+01}: 
\beq	
	{\tau_{LIF}} \frac{d V}{dt} = - (V-V_r) + I_{ds}(t) \nonumber \\ 
	\eeq
where $\tau_{LIF}$ is the membrane time constant and $V_r = -60$ mV is the rest voltage.  Spikes are produced when the voltage $V$ crosses $V_{thresh} = -45$ mV, at which point V is reset to $V_r$.  

For the parameters we have chosen, the downstream neuron is driven sub-threshold, so that $\langle  I_{ds}(t) \rangle$ is not sufficient to excite a spike --- any spikes must be driven by  {\it fluctuations} in $ I_{ds}(t)$. Thus, the variance of fluctuations in $V(t)$ should give
a rough estimate of how often membrane voltage will exceed the threshold, and consequently the downstream firing rate.  This variance is easy to compute for a passive membrane (i.e., neglecting spike-reset dynamics).  First, note that 
\begin{eqnarray}
V(t) & = & V_r + \int_{-\infty}^t I_{ds}(s) L(t-s) \, ds \nonumber
\end{eqnarray}
where $L$ is a one-sided exponential filter
\begin{eqnarray}
L(t) & = & \frac{1}{\tau} \exp (-t/\tau_{LIF}), \qquad t \ge 0 \nonumber \\
& = & 0,  \qquad t < 0. \nonumber
\end{eqnarray}
We compute the variance as follows, using the causality of $L$ to take each upper limit of integration to infinity:
\begin{eqnarray}
\Ex[V(t)^2] & = & \Ex \left[  \int_{-\infty}^t  (I_{ds}(s) - \langle I_{ds} \rangle) L(t-s) \, ds \int_{-\infty}^t (I_{ds}(r) - \langle I_{ds} \rangle) L(t-r) \, dr  \right] \nonumber \\
 & = &  \int_{-\infty}^{\infty} \int_{-\infty}^{\infty} ds \, dr \, L(t-s) L(t-r) \Ex \left[I(s) I(r) \right] \nonumber \\
 & = &  \int_{-\infty}^{\infty} \int_{-\infty}^{\infty} ds \, dr \, L(t-s) L(t-r) A_{ds}(s-r)  \nonumber  \\
%Ex[V(t)^2] & = &  \int_{-\infty}^{\infty}  dz \, A(z) \int_{-\infty}^{\infty} dr L(t-r-z) L(t-r)\\
%& = &  \int_{-\infty}^{\infty}  dz \, A(z) \int_{-\infty}^{\infty} dr L(-r-z) L(-r)   \\
& = &  \int_{-\infty}^{\infty}  dz \, A_{ds}(z) (L \ast \tilde{L})(-z)  \label{eqn:voltage_power_final}
\end{eqnarray}
where $\tilde{L}(t) \equiv L(-t)$; the last step involved the substitution $z=s-r$ and switching the order of integration.  This final interior integral can be evaluated in the Fourier domain: using the properties that $\mathcal{F}[\tilde{L}](\omega) = \mathcal{F}[L](-\omega)$ and the fact that for real functions $\mathcal{F}[f](-\omega) = \overline{\mathcal{F}[f](\omega)}$, we find
\begin{eqnarray}
\mathcal{F} \left[ (L \ast \tilde{L})\right](\omega) & = &  | \mathcal{F}[L](\omega)|^2  \nonumber \\
& = &  \frac{1}{1 + \omega^2 \tau_{LIF} ^2} \,.  \nonumber 
\end{eqnarray}
Therefore (for example by consulting a transform table)
\begin{eqnarray}
 (L \ast \tilde{L}) (t) & = & \frac{1}{2\tau_{LIF}} \exp(-|t|/\tau_{LIF}) \,. \nonumber 
 \end{eqnarray}
Substituting into Equation~\eqref{eqn:voltage_power_final}, we see that the variance of the downstream cell's voltage is given by a formula similar to that for the spike count covariances (Equation \ref{eqn:Cov_from_C}): both involve integrating the cross-covariance function against a (roughly) triangular-shaped kernel, with time scale $\tau_{LIF}$ in the former case and $T$ in the latter.   

Figure~\ref{fig:LIF_spike_rate}(c) shows the by-now familiar trends that this predicts.  For short membrane time scales $\tau_{LIF}$, Type II populations drive greater voltage variance; this is precisely analogous to the finding that spike-count correlations are greater for Type II cells over short time scales $T$.  For long $\tau_{LIF}$, Type I populations drive greater voltage variance, just as Type I spike trains are more correlated over long time scales $T$.  In panel (d), we compare this trend with actual firing rates elicited in the downstream cell (from numerical simulation).  The general trends match, validating our simple prediction.  

In sum, downstream neurons with short membrane time scales ($\tau_{LIF} \lsim 5$ ms) are preferentially driven by Type II cells upstream; for longer time scales, the preference shifts to Type I cells.  Some implications of this finding are noted in the Discussion.

\section*{DISCUSSION}

Diverging connections, leading to overlapped input shared across multiple neurons, are a
ubiquitous feature of neural anatomy.  We study the interplay between this connectivity pattern and basic properties of spike generation in creating collective spiking across multiple neurons.  We range spike generation over the fundamental categories of Type I to Type II excitability~\cite{excit,hodgkin48}.  The transition in excitability is produced by varying the A-current conductance $g_A$ within the well-studied Connor-Stevens neuron model.  

Our principal finding is that excitability type plays a major role in how shared --- i.e., correlated --- input currents are transformed into correlated output spikes.  Moreover, these differences depend strongly on the time scale $T$ over which correlations are assessed.  At short time scales $T$, Type-II neurons tend to produce relatively stronger spike correlations for comparable input currents~\cite{marella08,galan}.   At longer time scales, the opposite is generally true:   for a broad range of input currents, Type-I neurons transfer most of the shared variance in their inputs ($\sim 80\%$) into shared variance in output spikes, while Type-II neurons transfer less than half ($\sim 40\%$).  

We show that these results have direct implications for how downstream neurons with different membrane time constants will respond to Type I vs. Type II populations.  Specifically, downstream neurons preferentially respond to populations that are  strongly correlated on time scales similar to their membrane time constant.  Interestingly, for the case we study, we find that the breakpoint between selectivity to Type I vs. Type II populations was for downstream membrane time constants of $\approx 5$ ms, easily within the ranges found experimentally.

This raises interesting possibilities for neuromodulation.  The membrane time constant of the downstream cell could be changed by shunting affects of additional background inputs --- leading to a switch in its sensitivity to different upstream populations.  Alternatively, modulators applied to the upstream populations themselves could  change their excitability from Type I to Type II \cite{stiefel08,stiefel09}, adjusting their impact on a downstream cell with a fixed membrane time constant.

Overall, we demonstrate and apply a general principle:  the presence and balance among different membrane currents controls not only single-cell dynamics but also the strength and time scales of spike correlations in cell groups receiving common inputs.   We show how this relationship can be understood.  As a membrane current (here, $g_A$) is adjusted, firing rate - current curves progressively transition (here, from Type I to Type II).  At the same time, there is a transition in periodic orbit types that neural trajectories visit (here, ranging from orbits ``near" a fixed point to relatively ``isolated" orbits~\cite{rush95}).  In turn, this produces a steady progression of spike-triggered averages --- and hence the filters that neurons apply to shared input signals (here, from primarily integrating to primarily differentiating modes~\cite{ms08}, cf.~\cite{ayafb03}).  Basic formulas can then be used to translate these filtering properties into predictions for correlated spiking in neural pairs and populations~\cite{ostojic09} as well as the downstream impact of this cooperative activity.  We anticipate that this approach will bear fruit in studies of the collective activity of a wide variety of neuron types.

\subsubsection*{Relationship with prior work}

A number of prior studies have considered the problem of how spike generating dynamics affect the transfer of incoming current correlations into outgoing spike correlations~\cite{Bin+01,RochaDoironSJR07,SBJRD07,RosenbaumJ11,Tch+10,marella08,Vilela:2009p370,BarreiroST10,ostojic09,Tch+10,hong}.  In particular,~\cite{RochaDoironSJR07,SBJRD07,RosenbaumJ11} show that leaky integrate-and-fire (LIF) neurons can transfer up to 100\% of current correlations into spike count correlations.  The level transferred increases with the firing rate at which single neurons are operating, and the time scale $T$.  These findings are simpler to state compared with the present results for conductance-based neuron models, for which 100\% correlation transfer is never obtained, and trends with $T$ differ depending on $g_A$. 

Other works~\cite{hong,SBJRD07,Vilela:2009p370} investigate correlation transfer in more complex spiking models.  In particular,~\citetext{SBJRD07} and \citetext{Vilela:2009p370} explore the full parameter space of input currents for the quadratic integrate and fire model --- arguably that with the next level of complexity beyond the LIF model.  These authors find similar trends in correlation transfer as the neurons' operating points change, but a limitation to 66\% rather than $100 \%$ in correlation transfer.  Meanwhile, \citetext{hong} show complex dependencies on neural operating point for the Hodgkin-Huxley model.  Taken together, these studies suggested that correlation transfer depends on spike generating dynamics in a rich and diverse ways.  

This opened the door to a broader study, but exploring correlation transfer for the full space of possible spike generating dynamics in neural models is a daunting task.  The axis that spans from Type I to Type II excitability provides a natural focus.  This has been explored using sinusoidal ``normal form," phase-reduced models ~\cite{marella08,galan,BarreiroST10,abouzeidArXiv}.  These studies used simulations in the superthreshold regime, together with analysis in the limits of very short or very long time scales $T$, to show the same trend in correlation transfer over short vs. long $T$ that we find here for conductance-based models.  A greater frequency of instantaneous (small $T$) spikes for Type II vs. Type I neurons was predicted using these simplified models~\cite{marella08,galan}; later, \citetext{BarreiroST10} predicted the switch in relative correlation transfer efficiency from Type II to Type I models as $T$ increases.

The present contribution is to test the resulting predictions using biophysical, conductance-based models valid in a wider range of firing regimes, to explain the origins of variable correlation transfer via filtering properties of Type I vs. Type II cells, and to demonstrate the impact on downstream neurons.

\subsubsection*{Scope, limitations, and open questions}

The circuit model that we have studied, as illustrated in Fig.~\ref{fig:schematic}, is limited to a single, idealized feature of feedforward connectivity:  overlapping inputs to multiple recipient cells.  More realistic architecture could include delays in incoming inhibitory vs. excitatory inputs~\cite{Gabernet+05}.  Interactions of shared-input circuitry with recurrent connectivity  also pose important questions~\cite{CHENGANDERM}.  This is especially so given the distinct properties of Type I vs. Type II cells in synchronization due to reciprocal coupling~\cite{excit,ErmentroutTBook}.

Other aspects of our biophysical and circuit dynamics are also idealized.  For one, individual input currents fluctuated on arbitrarily fast time scales (i.e., as white noise processes).  Relaxing this would be an interesting extension.  While prior studies~\cite{RochaDoironSJR07} suggest that trends will persist for inputs with fast (but finite) time scales, new effects could arise for slower-time scale inputs representative of slower synapses or even network-level oscillations.
Another addition would be for inputs to arrive via excitatory and inhibitory conductances, rather than currents; while previous studies with integrate and fire cells~\cite{RochaDoironSJR07} have found that this yields qualitatively similar results, there could be interesting interactions with underling filtering properties in biophysical models.  The same holds true for inputs that arrive at dendrites in multi-compartment models.  

Likewise, the circuitry of Fig.~\ref{fig:LIF_spike_rate}(a) that we used to investigate the impact of correlated spiking on downstream neurons was highly idealized.  An especially appealing extension would be to note that inhibitory and excitatory neurons often have different excitability types.  Thus, downstream cells could receive input from both excitatory Type I and inhibitory Type II populations.  Our results suggest that sensitivity to excitatory vs. inhibitory afferents would vary with membrane time constants downstream, possibly amplifying the modulatory effects identified here.

%\bibliography{corr_transfer_CS,corr_transfer_CS_ESB_adds}

% Included for ArXiv

\end{document}